\documentclass[preprint,12pt]{elsarticle}
\usepackage{placeins} 
\usepackage[utf8]{inputenc}
\usepackage[T1]{fontenc}

\usepackage{graphicx} 
\usepackage{comment}
\usepackage{natbib}
\usepackage{tikz} 
\usepackage{pgfplots} \pgfplotsset{compat=newest}
\usepackage{hyperref}
\usepackage{graphicx}	
\usepackage{amsmath}	
\usepackage{amssymb}	
\usepackage{multicol}   
\usepackage{bm}		    
\usepackage{pdflscape}	
\usepackage{soul}       
\usepackage[normalem]{ulem} 
\usepackage{xcolor}     
\usepackage{array}      
\usepackage{caption}    
\usepackage{cleveref}   
\usepackage{float}
\usepackage{aas_macros}

\usepackage{xcolor} 
\definecolor{FelipeColor}{RGB}{204,0,0} 

\newif\iffelipemark
\felipemarktrue   

\begin{document}

\begin{frontmatter}

\title{Cosmological dynamics and observational constraints of an interacting early scalar field coupled to radiation}

\author[inst1]{Dorian Araya \corref{cor1}}
\ead{dorian.araya.a@mail.pucv.cl}

\author[inst1]{Felipe Herrera}
\ead{felipe.herrera.f01@mail.pucv.cl}

\author[inst1]{Nelson Videla}
\ead{nelson.videla@pucv.cl}

\cortext[cor1]{Corresponding author}

\affiliation[inst1]{
  organization={Instituto de F\'isica, Pontificia Universidad Cat\'olica de Valpara\'iso, Avenida Universidad 330},
  city={Valpara\'iso},
  country={Chile}
}

\begin{abstract}
We study the cosmic evolution of an interacting early scalar 
field–radiation model, in which a minimally coupled scalar field exchanges energy with the radiation sector through an exponential coupling. Extending previous formulations, a non-relativistic matter component is included explicitly, which allows a self consistent description of cosmological dynamics from the radiation-dominated era to late-time acceleration. Analytical expressions for the background expansion are derived and characterized using kinematic diagnostics. We constrain the model using observational Hubble data, Type Ia Supernovae, baryon acoustic oscillations (including DESI DR2), and compressed cosmic microwave background distance information, performing a Bayesian MCMC analysis. The interaction parameter is found to be consistent with zero, though small deviations from standard radiation scaling are allowed. These deviations can partially alleviate the Hubble tension by modifying the sound horizon, but this is accompanied by correlated shifts in the matter density. The reconstructed expansion history remains close to $\Lambda$CDM at late times. Model comparison suggest that the interacting scenario is statistically competitive but not decisively preferred by current background data.
\end{abstract}


\end{frontmatter}

\section{Introduction}\label{Introduction}

Dark energy (DE) is perhaps the most intriguing component of the Universe, accounting for nearly $70\%$ of its total energy density and driving the observed late-time cosmic acceleration as confirmed by current observational evidence coming from precise measurements of Supernovae Ia (SnIa) \cite{SupernovaSearchTeam:1998fmf,SupernovaCosmologyProject:1998vns,SupernovaCosmologyProject:2008ojh}, cosmic microwave background (CMB) anisotropies,\cite{WMAP:2010sfg,WMAP:2010qai,Planck:2015bue,Planck:2018vyg}, and  baryon acoustic oscillations (BAO) \cite{BOSS:2012xge,Blake:2011rj,SDSS:2009ocz,SDSS:2003eyi,2dFGRS:2005yhx}. In particular, within the $\Lambda$CDM model, DE is described by a cosmological constant (CC) \cite{Padmanabhan:2002ji,Sahni:1999gb,Carroll:2000fy}. At the same time, the remaining content is dominated by cold dark matter (DM) ($\sim 25\%$), baryonic matter ($\sim5\%$), and a small contribution from relativistic species (photons and neutrinos) \cite{Planck:2018vyg}. Despite the $\Lambda$CDM cosmology being remarkably successful for a wide range of cosmological observations \cite{Bull:2015stt}, it still faces several theoretical and observational challenges, namely: (i) the cosmological constant problem \cite{Weinberg:1988cp,Martin:2012bt},  (ii) the cosmic coincidence problem (or the so-called “why now”
problem) \cite{Zlatev1998,Arkani-Hamed:2000ifx}, and more recently, (iii) the tension between measurements of the Hubble parameter $H_0$  \cite{DiValentino:2021izs,Riess:2020fzl}, and (iv) the $S_8$ tension ($S_8=\sigma_8 \sqrt{ \Omega_m/0.3}$)   \cite{DiValentino:2020vvd}. Theoretically, tensions between measurements in the early and the late universe could represent the signature of new physics beyond the $\Lambda$CDM model \cite{Mortsell:2018mfj}. In particular, discrepancies in the Hubble constant $H_0$ observed by comparing early CMB measurements 
\cite{Planck:2018vyg} with late local distance-ladder observations 
\cite{Riess2018,Riess2019}, as well as deviations in the growth rate of cosmic structures and in 
weak lensing measurements 
\cite{Dossett2012GrowthConstraints,DESY3_2022,Asgari2021_KiDS1000,Troester2020_S8}, 
have motivated scenarios that extend the standard cosmological model; see, e.g., \cite{DiValentino2016Extensions,Efstathiou:2024dvn}. These include, for instance, dynamical DE models with a varying equation of state (EoS) parameter, as suggested by BAO measurements from the first data release of the Dark Energy Spectroscopic Instrument (DESI) \cite{DESI:2024mwx}, with the recent second data release (DR2) \cite{desi2025} in combination with CMB observations and various SnIa catalogs. Another promising proposal is the inclusion of an early dark energy (EDE) component, whose contribution to the Universe’s total energy content is non-negligible around the epoch of matter–radiation equality or recombination \cite{Doran:2006}. Subsequently, it dilutes faster than matter or radiation, becoming negligible at late times \cite{Wetterich:2004}. Furthermore, a scalar–radiation Interaction can partially alleviate the Hubble tension by reducing the sound horizon at last scattering through an enhanced expansion rate before recombination \cite{Karwal:2016, Poulin:2018}. The realization of EDE is usually achieved through scalar fields with dynamics that briefly become relevant in the early Universe, distinguishing this mechanism from the constant vacuum energy of the standard $\Lambda$CDM model \cite{Smith:2020, Hill:2020}. In relation to the observational constraints on EDE models, they have been confronted by CMB, BAO, SnIa, and LSS data, yielding that any EDE contribution to the total energy density near recombination must be small, typically $f\lesssim 0.07-0.1$ (95$\%$ C.L.), to avoid spoiling the excellent fit of $\Lambda$CDM to the CMB and galaxy clustering \cite{Poulin:2023lkg,Qu:2024lpx}. Despite large values of EDE not being supported by current observations, a small EDE fraction can partially alleviate the Hubble tension by increasing the inferred value of $H_0$ from early-Universe data \cite{Poulin:2025nfb}. In this respect, the whole agreement with local measurements is not achieved when all the data are included (CMB + BAO + LSS) \cite{Poulin:2025nfb}. The $S_8$ tension is a serious challenge for EDE models, since they generally predict enhanced clustering amplitudes that worsen the disagreement with low-$S_8$ LSS data \cite{Hill:2020osr}. Recently, several extensions of the standard EDE scenario have been explored to alleviate tensions with LSS data while being compatible with CMB and BAO constraints. For instance, a coupling between EDE and DM allows for suppressing structure growth and lowering the predicted $S_8$ amplitude \citep{Yashiki:2025loj}. These approaches aim to improve cosmological concordance but have yet to fully resolve both the $H_0$ and $S_8$ tensions simultaneously. A recent proposal along these lines was introduced in Ref.\cite{Bisabr_2025}, where an interacting EDE model is proposed, in which energy exchange occurs between EDE and radiation rather than between EDE and DM. In particular, EDE comes in 
the form of a minimally coupled scalar field, which has a direct coupling with the radiation sector during the radiation-dominated epoch. The considered action in \cite{Bisabr_2025} is given by
\begin{equation}\label{eq:0}
S \;=\; \int d^4x\,\sqrt{-g}\,\Big\{
\tfrac{1}{16\pi G}R \;-\; \tfrac{1}{2}\,g^{\mu\nu}\nabla_\mu\phi\nabla_\nu\phi \;-\; V(\phi)
\;\;+\; C(\phi)\,\mathcal{L}_m
\Big\},
\end{equation}
where $R$ is the Ricci scalar, while $g$ is the determinant of the metric tensor $g_{\mu \nu}$. As can be seen from the above action, the scalar field $\phi$ with potential $V(\phi)$ interacts directly with the matter field (represented by the Lagrangian density $\mathcal{L}_m$ ) through the coupling function 
$C(\phi)$. This setup naturally emerges in theories such as Brans–Dicke and $f(R)$ gravity after
perform a conformal transformation \cite{Sotiriou:2008rp,DeFelice:2010aj}, as well as in chameleon models \cite{Khoury:2003aq}, where the scalar field gets a coupling to the matter Lagrangian by means of a function $C(\phi)$. To account for a coupling between the scalar field and radiation, the only matter field present is radiation, i.e. $\mathcal{L}_m =\mathcal{L}_{\gamma}$, where $\gamma$ denotes radiation. In particular, for an exponential form of the coupling $C(\phi)=e^{-\sigma \phi}$, with $\sigma$ being a coupling constant, the energy density of radiation becomes modified as $\rho_{\gamma}\propto a^{-4+\epsilon}$. In the latter expression, $\epsilon\equiv \frac{4\sigma \phi}{3 \ln{a}}$ represents the energy exchange between the scalar field and radiation, and it is assumed to be constant. Such a scalar-photon coupling  provides a minimal and physically well-motivated mechanism to modify the pre-recombination expansion history, directly impacting the sound horizon while the cosmological dynamics at late-times is not altered \cite{Bisabr_2025,Buen-Abad:2025bgd}. In addition, the evolution of CMB temperature with redshift is modified as
\begin{equation}
    T(z)=T_0(1+z)^{1-\beta},\label{cmbt}
\end{equation}
where $\beta=\epsilon/4$. Similar modifications to the standard adiabatic law $T(z)=T_0(1+z)$ have already been found in \cite{Avgoustidis:2011aa,Avgoustidis:2013bqa}, where phenomenological scenarios involving scalar fields coupled to radiation were studied. It was shown in \cite{Avgoustidis:2013bqa} that, using the most updated datasets available at this time, $\beta$ was constrained to be $\beta=0.004\pm 0.016$ up to a redshift $z \sim 3$. Current direct measurements of the CMB temperature–redshift relation
from thermal Sunyaev–Zel’dovich (SZ) measurements in clusters, together with molecular/atomic excitation constraints up
$z\simeq 6$, the most recent combined analysis finds $\beta=-0.0106\pm 0.0124$ (68\% C.L.) \cite{Ruchika2025}, consistent with the standard adiabatic scaling. On the other hand, model-dependent analyses that include distance-duality tests further constrain $|\beta|\lesssim 10^{-3}$ \cite{Avgoustidis2016}, leaving no significant evidence for departures from the standard thermal history. If an explicit non-relativistic matter sector $ \mathcal{L}_{m}$ is incorporated into the action (\ref{eq:0}), a more realistic description of the Universe’s dynamics across the radiation–matter transition will be provided, allowing the evolution of the coupled scalar field through recombination and the onset of cosmic acceleration. In this way, the main goal of the present work is to extend the interacting EDE scenario proposed in \cite{Bisabr_2025} by introducing an explicit matter component into the action, thereby generalizing the coupling between the scalar field and the radiation sector to a cosmologically complete framework. In this sense, we will investigate how the inclusion of non-relativistic matter affects the background dynamics, the effective equation of state, and the overall expansion history of the Universe. Furthermore, the resulting interacting scalar field-radiation model will be tested against recent cosmological observations — including  observational Hubble data (OHD), CMB, BAO, and SnIa data — through a Markov Chain Monte Carlo (MCMC) analysis. We also consider two priors for $\Omega_{m0}$, namely: (i) a fully free prior (\textbf{CI}), and (ii) a Planck-motivated Gaussian prior (\textbf{CII}), to evaluate the robustness of the inferred interaction and its impact on $H_0$.

Our work is organized as follows: after this introduction, in Section \ref{Cosmological Model}, we develop the theoretical framework for the interacting scalar-field-radiation system. In particular, the main background-level equations describing cosmological evolution are derived. This includes obtaining an analytical solution for the Hubble rate $H(z)$ as a function of redshift and introducing the statefinder parameters. In Section \ref{Cosmological Data}, we present the data and relevant equations for implementing the MCMC analysis. In Section \ref{Results}, we present the results of the MCMC analysis and numerically evaluate the evolution of the universe at redshift, comparing it with that of the $\Lambda$CDM model. Finally, we summarize our findings and present our conclusions in Section \ref{Discussion and Conclusions}. Throughout our work, we adopt the mostly positive metric signature $(-,+,+,+)$ and use natural units, with $c=\hbar=1$. \\

\section{Scalar–Radiation Interaction: Theoretical Setup}\label{Cosmological Model}

Our starting point is a consistent modification of the original action, as proposed in \cite{Bisabr_2025}, by adding a matter term in the action without violating observational bounds on fifth forces or equivalence-principle tests \cite{Burrage2018,Brax2021}
\begin{equation}\label{eq:1}
S \;=\; \int d^4x\,\sqrt{-g}\,\Big\{
\tfrac{1}{16\pi G}R \;-\; \tfrac{1}{2}\,g^{\mu\nu}\nabla_\mu\phi\nabla_\nu\phi \;-\; V(\phi)
\;\;+\; \mathcal{L}_{m} \;+\; C(\phi)\,\mathcal{L}_\gamma
\Big\}.
\end{equation}

Where $\mathcal{L}_m$ and $\mathcal{L}_\gamma$ are the Lagrangian densities of non-relativistic matter (cold dark matter plus baryonic matter) and radiation, respectively. Matter is minimally coupled, while the radiation sector couples conformally to the scalar field through the function $C(\phi)=e^{-\sigma\phi}$, where the parameter $\sigma$ represents the strength of the interaction. 
Varying action (\ref{eq:1}) with respect to the metric $g_{\mu \nu}$ and the scalar field $\phi$ yields, respectively
\begin{align}
    G_{\mu \nu}=8\pi G T_{\mu \nu}&=8\pi G\left(T_{\mu \nu}^{(m)}+C(\phi)T_{\mu \nu}^{(\gamma)}+T_{\mu \nu}^{(\phi)}\right),\label{EC}\\
\Box\phi - V_{,\phi}(\phi)  &= -C_{,\phi}(\phi)\,\mathcal{L}_\gamma,\label{KG}   
\end{align}
were $_{,\phi}$ denotes derivatives with respect to $\phi$. $T_{\mu \nu}^{(\phi)}$ represents the energy-momentum tensor for the scalar field, given by
\begin{equation}
    T_{\mu \nu}^{(\phi)}=\left(\nabla_{\mu}\phi \nabla_{\nu}\phi-\frac{1}{2}g_{\mu \nu}\nabla_{\alpha}\phi \nabla^{\alpha}\phi\right)-V(\phi)g_{\mu \nu},\label{Tmu nu phi}
\end{equation}
while 
\begin{equation}
    T_{\mu \nu}^{(i)}=\frac{-2}{\sqrt{-g}}\frac{\delta\left(\sqrt{-g} \mathcal{L}_i\right)}{\delta g^{\mu \nu}},
\end{equation}
represents the energy-momentum tensor for non-relativistic matter or radiation fields ($i=m,\gamma$).
Besides, considering that the total energy–momentum tensor is covariantly conserved i.e., $\nabla^{\mu}T_{\mu \nu}=0$, and that individual components need not be conserved, from the right side of Eq.(\ref{EC}) we obtain
\begin{equation}
 \nabla^{\mu} T^{(\phi)}_{\mu\nu}
+ \nabla^{\mu}\!\left[ C(\phi)\, T^{(\gamma)}_{\mu\nu} \right]
+ \nabla^{\mu} T^{(m)}_{\mu\nu}
= 0.
\end{equation}
Because matter is minimally coupled
\begin{equation}
    \nabla^{\mu}T_{\mu \nu}^{(m)}=0.\label{consm}
\end{equation}
Accordingly, the interaction is solely between radiation and the scalar field
\begin{equation}
    \nabla^{\mu} T^{(\phi)}_{\mu\nu}
= -\,\nabla^{\mu}\!\left[\,C(\phi)\, T^{(\gamma)}_{\mu\nu}\right].\label{coupled}
\end{equation}
Expanding the right-hand side of Eq.(\ref{coupled})
\begin{equation}
 \nabla^{\mu}\left[C(\phi)T^{(\gamma)}_{\mu \nu}\right]=C_{,\phi}(\phi)\nabla^{\mu}\phi \,T^{(\gamma)}_{\mu \nu}+C(\phi)\nabla^{\mu}T^{(\gamma)}_{\mu \nu}.
\end{equation}
Radiation is not separately conserved due to coupling. By using KG equation (\ref{KG}), the coupling $C(\phi)=e^{-\sigma\phi}$, and assuming that the Lagrangian density has the form $\mathcal{L}_{\gamma}=p_{\gamma}=\rho_{\gamma}/3$ \cite{Bisabr_2025}, with $p_{\gamma}$ and $\rho_{\gamma}$ being respectively the pressure and energy density of radiation, the time component of left-hand side of Eq.(\ref{coupled}) for a spatially flat FLRW  becomes
\begin{equation}
\dot\rho_\phi + 3H(1+w_\phi)\rho_\phi  = -\,\frac{\sigma}{3}\,e^{-\sigma\phi}\,\dot\phi\,\rho_\gamma, \label{eq:2}  
\end{equation}
which is the conservation for the scalar field in the
perfect fluid form with a source term. Here $w_{\phi}=\frac{p_{\phi}}{\rho_{\phi}}$ is the equation of state (EoS) of the scalar field, with $\rho_{\phi}=\frac{\dot{\phi}^2}{2}+V(\phi)$ and $p_{\phi}=\frac{\dot{\phi}^2}{2}-V(\phi)$. Now, upon replacement of Eq.(\ref{eq:2}) into (\ref{coupled}), the modified conservation equation for radiations gives
\begin{equation}
    \dot\rho_\gamma + 4H\,\rho_\gamma = \frac{4}{3}\,\sigma\,\dot\phi\,\rho_\gamma.\label{eq:3}
\end{equation}
In addition, the conservation equation for non-relativistic matter has the usual form
\begin{equation}
\dot\rho_{m} + 3H\,\rho_m=0.\label{ccm}    
\end{equation}

Conversely, the following set of equations is satisfied by the Hubble rate
\begin{align}
3H^2 &= 8\pi G \left(e^{-\sigma \phi}\rho_{\gamma}+\rho_{\phi}+\rho_m\right),\label{HF}\\
2\dot{H} +3H^2&= -8\pi G\left(e^{\sigma \phi}p_{\gamma}+p_{\phi}\right). \label{dot}
\end{align}

Eq.(\ref{eq:3}) can be integrated for $\rho_{\gamma}$, yielding

\begin{equation}\label{eq:5}
\rho_\gamma(a) \;=\; \rho_{\gamma 0}\,a^{-4+\epsilon}.
\end{equation}
Here, $\rho_{\gamma 0}$ represents the energy density of radiation evaluated at $z=0$, while $\epsilon$ is defined as
\begin{equation}\label{eq:6}
\epsilon \;\equiv\; \frac{4\sigma\phi}{3 \ln a}.
\end{equation}
This latter quantity quantifies the energy transfer between the scalar field and radiation. If $\epsilon >0$, energy flux goes from the scalar field to radiation, which implies the latter dilutes more slowly than in the standard case $\rho_{\gamma}\propto a^{-4}$.  In contrast, for the case $\epsilon<0$, energy is injected into the scalar field, yielding a faster dilution of $\rho_{\gamma}$ and consequently decreasing the sound horizon. Under the assumption that both $w_\phi$ and $\epsilon$ are constant, it is found that the scalar field has a logarithmic dependence on the scale factor as
\begin{equation}
 \phi(a)=\gamma\ln a ,\label{phia} 
\end{equation}
where $\displaystyle \gamma=\tfrac{3\epsilon}{4\sigma}$. The assumption of treating the EoS parameter of the scalar field $w_{\phi}$ and the interaction parameter $\epsilon$ as constants should be understood as an effective description of the background dynamics. In interacting scalar-field models with exponential couplings and/or potentials, $w_{\phi}$ and logarithmic evolution $\phi \propto \ln a$ naturally arise \cite{Ferreira:1997hj,Copeland:1997et,Amendola:1999er}. In addition, the assumption of $\epsilon$ being constant captures the radiation-dominated and early matter-dominated epochs. Deviations at very late times are negligible because radiation becomes dynamically irrelevant.

During Big Bang Nucleosynthesis (BBN), the modified scaling of the radiation (\ref{eq:5}) leads to a fractional change in the energy density of radiation $\delta \rho_{\gamma}/\rho_{\gamma}\simeq \epsilon \ln (1+z_{BBN})$, where $z_{BBN}\sim 10^9$. Such a modification can be expressed as an effective change in the number of relativistic degrees of freedom, $\Delta N_{\text{eff}}$, through $\delta \rho_{\gamma}/\rho_{\gamma}\simeq 0.134 \Delta N_{\text{eff}}$. Current BBN bounds on $\Delta N_{\text{eff}}$, $|\Delta N_{\text{eff}}| \lesssim 0.4$ (95\% C.L.) \cite{Steigman:2012ve}, which requires that $|\epsilon|\lesssim \mathcal{O}(10^{-2})$. 

Upon replacement of Eqs. (\ref{eq:5}) and (\ref{eq:6}) into (\ref{eq:2}), the energy density of the scalar field can be found as follows
\begin{equation}\label{eq:8}
\rho_\phi(a)=a^{-3(1+w_\phi)}\!\left[\rho_{\phi0}-\frac{\epsilon}{4\lambda}\,\rho_{\gamma0}\,\big(a^\lambda-1\big)\right],
\end{equation}
where $\rho_{\phi0}$ denotes the density of the scalar field evaluated at $z=0$. For simplicity, it is introduced the parameter $
\lambda \equiv -1 + 3w_\phi + \frac{\epsilon}{4}$. In addition, the energy density of matter can be obtained from direct integration of Eq.(\ref{ccm}), giving

\begin{equation}\label{eq:7}
\rho_{m}(a) = \rho_{m0}\,a^{-3},
\end{equation}
with $\rho_{m 0}$ being the current value of the density of matter.

The Friedmann equation (\ref{HF}) can be rewritten by using Eqs.(\ref{eq:5}), (\ref{eq:8}), (\ref{eq:7}), and the relation between the scale factor $a$ and the redshift $z$, given by $a = (1 + z)^{-1}$. At the same time, diving by $H_0^2$ (the Hubble constant squared), it yields

\begin{equation}\label{eq:10}
\begin{aligned}
E(z)^2\equiv \frac{H(z)^2}{H_0^2}
&= \Omega_{m0}(1+z)^3
+\Big(1-\frac{\epsilon}{4\lambda}\Big)\Omega_{\gamma0}(1+z)^{4-\epsilon/4}\\
&\hspace{3em}
+\Big(\Omega_{\phi0}+\frac{\epsilon}{4\lambda}\Omega_{\gamma0}\Big)(1+z)^{3(1+w_\phi)}.
\end{aligned}
\end{equation}
 In (\ref{eq:10}), the dimensionless density parameter for each component $i$ is defined as $\Omega_{i0} = \rho_{i0}/\rho_{0\text{crit}}$, where $\rho_{0\text{crit}}$ is the critical density evaluated today defined as $\rho_{0\text{crit}} = 3H_0^2/8\pi G$. {\bfseries
While radiation depends on the reduced Hubble constant $h$, according to the standard relation
\begin{equation}
\boldsymbol{\Omega_{\gamma 0}}(h) = \boldsymbol{\Omega_{r0}}\left(1 + 0.2271\,\boldsymbol{N_{\rm eff}}\right),
\label{eq:Omega_r}
\end{equation}
where $\boldsymbol{\Omega_{\gamma 0}}$ denotes the present photon density parameter, given by
\begin{equation}
\boldsymbol{\Omega_{r0}} = 2.469 \times 10^{-5}\,h^{-2},
\label{eq:Omega_gamma}
\end{equation}
and $\boldsymbol{N_{\rm eff}}$ is the effective number of relativistic species
\cite{Neff_relavistic_species}. Throughout this work, we fix
$\boldsymbol{N_{\rm eff}} = 3.04$, corresponding to the standard cosmological scenario
with three relativistic neutrino species.
}

Imposing $E(z=0)=1$, we obtain the flatness condition

\begin{equation}\label{eq:11}
\Omega_{\phi 0}\ = 1 -\;\; \Omega_{m0} \;-\; \Omega_{\gamma 0}. \;
\end{equation}
Moreover, using equation \ref{eq:10}, we can define the time scale of the universe in function of redshift as
\begin{equation}\label{eq14}
 t(z)=\frac{1}{H_0} \int_z^{\infty} \frac{1}{(1 + z)\,E(z)}  dz,
\end{equation} 
where the time $t(z)$ is expressed in [Gyr].\\
Furthermore, we compute the effective EoS parameter $w_{\mathrm{eff}}(z)$ for our cosmological model, which encodes information about the universe's composition, the evolution of its energy density and the dynamics of its expansion. The general expression for $w_{eff}$ depends on the first derivative of $E^2(z)$ with respect to the redshift $z$ as follows
\begin{equation}
w_{\mathrm{eff}}(z)
= -1 + \frac{1}{3}\,\frac{d\ln E^{2}(z)}{d\ln(1+z)}
.
\end{equation}

In addition to $w_{\mathrm{eff}}(z)$, we introduce the deceleration parameter $q(z)$ in terms of derivatives of  (\ref{eq:10}) as
\begin{equation}\label{eq:12}
q(z) = \frac{(z+1)}{{E(z)}}\frac{dE(z)}{dz} - 1.
\end{equation}
Besides, it is defined the jerk parameter $j(z)$ in terms of derivatives of (\ref{eq:12})
\begin{equation}
  j(z) =\ q + 2q^{2} - (1+z)\frac{dq}{dz}.
\end{equation}
Finally, we define the parameter $s(z)$, which it should not to be confused with the snap parameter (fourth time derivative), given by:
\begin{equation}
s(z) = \frac{j - 1}{3\left(q - \tfrac{1}{2}\right)} .
\end{equation}

The set of parameters $\{q(z),j(z),s(z)\}$ states the basis of the statefinder diagnostic \cite{statefinder_paper}, which gives information to analyze how alternative cosmological models deviate from the $\Lambda$CDM model. In particular, on the $s$-$j$ plane, the fixed point $\{s,j\}=\{0,1\}$ identifies the $\Lambda$CDM model. The fixed point $\{s,j\}=\{1,1\}$ corresponds to SCDM, which is the Einstein de Sitter with $\Omega_m=1$, $\Omega_{\mathrm{de}}=0$. On the $q$-$j$ plane, the fixed point $\{q,j\}=\{-1,1\}$ represents a pure de Sitter universe (dS), corresponding to $\Omega_{\mathrm{de}}=1$, which acts as the asymptotic attractor of $\Lambda$CDM.

\section{Cosmological data}\label{Cosmological Data}
\label{constraintsdatasamples}
In this section, we present the cosmological data sets employed in the Bayesian analysis to constrain the interacting scalar–radiation model. Our analysis is restricted to background observables that probe the expansion history of the Universe. At the background level, the model is characterized by four free parameters: the reduced Hubble constant $h\equiv H_0/100$, the density parameter of matter at present time $\Omega_{m0}$, the interaction parameter $\epsilon$, and the EoS parameter of the scalar field $w_{\phi}$. 

 \subsection{ Observational Hubble Data}\label{sec:OHD}
Galaxies evolving passively over timescales much longer than their age differences can be used as cosmic chronometers (CC) to measure the Hubble parameter via the differential age method within the FLRW framework \cite{Jimenez_2002}. In practice, the idea is to observe two massive galaxies that share as closely as possible the same formation epoch, identified spectroscopically through the 4000\,\AA\ break, and that lie at very close redshifts. Their spectroscopic age difference $\Delta t$ and redshift gap $\Delta z$ provide a direct estimate of the Hubble parameter,
\begin{equation}
    H(z) = - \frac{1}{1+z}\,\frac{dz}{dt}.
\end{equation}

Where $dt$  is the differential time evolution
of the universe at redshift interval $dz$. Thus, we can constraint the cosmological model using observational Hubble data (OHD) samples \cite{DATAHZ, HZnewpoint1, hznewpoint2, hznewpoint3}. With this samples, we can use the chi-square function given by

\begin{equation}
\chi^2_{\text{OHD}} = \sum\limits_{i=1}^{N} \left( \frac{H_{\text{th}}(z_i) - H_{\text{obs}}(z_i)}{\sigma_{i,\text{obs}}} \right)^2,
\end{equation}
where $H_{\text{th}}(z_i)$ is the theoretical Hubble parameter  $H_{\text{obs}}(z_i) \pm \sigma_{i,\text{obs}}$ is the observational Hubble parameter  (from the differential age method) with its uncertainty at the redshift $z_i$, and $N$ is the number of points used. Where $N=33$ and the redshift range is $0.07<z<1.965$.

\subsection{Type Ia Supernovae}\label{sec:SNIa}
Type Ia supernovae (SnIa) arise in binary systems where a white dwarf accretes material from its companion, often hydrogen or helium \cite{whitedwarcompo}. As the white dwarf approaches the Chandrasekhar limit ($\sim 1.44\,M_\odot$), thermonuclear runaway of carbon and oxygen is triggered in the core \cite{chandrasekharlimit, biografiaSNIa}, producing an explosion with a peak luminosity that can be standardized. This makes SnIa powerful distance indicators. Samples of SnIa \cite{SNia_Riess_2018_2,SNia_Scolnic_2018_v2,SNIa_Magana:2017nfs} provide distance modulus measurements across a wide redshift range and are routinely used to constrain the background expansion. In this work we use the Pantheon+ compilation \cite{DATASN}, which contains 1701 events and reports the distance modulus $\mu$ together with the redshift $z$ measured in the CMB frame. The dataset is calibrated using the Cepheid distance scale and includes the SH0ES Cepheid host distance anchors.

To account for correlated measurements and to marginalize analytically over the absolute magnitude, we adopt the standard form of the marginalized chi–square,
\begin{equation}\label{eq:chi_snia}
 \chi^2_{\text{SNIa}} = \alpha + \log\!\left(\frac{\gamma}{2\pi}\right) - \frac{\beta^2}{\gamma},
\end{equation}
with
\[
\alpha = (\Delta{\mu})^T \cdot \text{Cov}_P^{-1} \cdot \Delta{\mu},\qquad
\beta = (\Delta{\mu})^T \cdot \text{Cov}_P^{-1} \cdot \Delta1,\qquad
\gamma = \Delta1^T \cdot \text{Cov}_P^{-1} \cdot \Delta1,
\]
where $\Delta\mu$ is the vector of residuals between theoretical and observed distance moduli, $\Delta1=(1,1,\ldots,1)^T$, and the total covariance is $\text{Cov}_P=\text{Cov}_{P,\text{sys}}+\text{Cov}_{P,\text{stat}}$.

The theoretical distance modulus is estimated by
\begin{equation}
m_{\text{th}} = M + 5 \log_{10} \left[\frac{D_L(z)}{10 \text{pc}}\right],
\end{equation}
\noindent
where $ M $ is a nuisance parameter which has been marginalized in \eqref{eq:chi_snia}
and the luminosity distance ($D_{L}$) is given by

\begin{equation}
D_L(z) = (1+z) \cdot \frac{c}{H_0} \chi(z),
\end{equation}

being $\chi(z)$ the comoving radial distance given by:

\begin{equation}
    \chi(z) = \int_0^{z} \frac{dz'}{E(z')}. \qquad
\end{equation}
Thus, the Pantheon+ redshift coverage $0.01<z<2.5$ enables late–time expansion constraints.

\subsection{Baryonic Acoustic Oscillations}
BAO emerged from the interplay of two fundamental forces during the universe's early stages: gravity and pressure. These forces clashed in an immensely dense region of primordial plasma composed mainly of electrons and baryons (protons and neutrons). The plasma confined photons through Thomson scattering, exerting significant outward pressure, while simultaneously, gravity pulled matter towards the dense center.

As the universe expanded and cooled during the epoch of recombination $(z\simeq 1100)$, hydrogen atoms began to form from baryonic matter. With hydrogen being electrically neutral, photons could travel more freely through space, no longer trapped by constant interactions. This allowed the photons to decouple from the plasma, leaving behind patterns of baryonic matter in the form of waves.

These patterns, or BAO, leave an imprint on the CMB. The distance from the central dense region to the first ripple of baryonic matter, where galaxies later formed, became a standard ruler for measuring cosmic expansion \cite{bassett2009baryon}. 
The theoretical angular scale ($\theta_{\text{th}}$) is estimated as

\begin{equation}
\theta_{\text{th}}(z) = \frac{r_{\text{drag}}}{(1 + z)D_{\text{A}}(z)}.
\end{equation}

The comoving sound horizon, $r_{s}(z)$, is defined as

\begin{equation}\label{eq:24}
r_{s}(z) = \frac{c}{H_{0}} \int_{0}^{z} \frac{c_{s}(z^{'})}{E(z^{'})} \, dz^{'},
\end{equation}
where the sound speed $c_{s}(z) = \frac{1}{\sqrt{3(1 + R_{\bar{b}}/(1 + z))}}$, with \newline $R_{\bar{b}} = 31500 \Omega_{b}h^{2}(T_{\text{CMB}}/2.7\,\text{K})^{-4}$, and $T_{\text{CMB}}$ is the CMB temperature. The redshift $z_{\text{drag}}$ at the baryon drag epoch is well-fitted with the formula proposed by \cite{BAOFORMULA} given by
\begin{equation}
\begin{aligned}
z_{\text{drag}} = & \ 1291(\Omega_{m0}h^{2})^{0.251} \left[1 + 0.659 (\Omega_{m0}h^{2})^{0.828} \right. \\
& \hspace{3cm} \cdot \left. \left(1 + b_{1}\left(\frac{\Omega_{b0}h^{2}}{b_{2}}\right)\right)\right],
\end{aligned}
\end{equation}
where
\begin{align*}
b_{1} &= 0.313 (\Omega_{m0} h^{2})^{-0.419} h^{1 + 0.607(\Omega_{m0} h^{2})^{0.674}}, \\
b_{2} &= 0.238 (\Omega_{m0} h^{2}).
\end{align*}
where $\Omega_{\text{m}0}=\Omega_{dm0}+\Omega_{\text{b}0}$ at $z = 0$ respectively. Although the radiation evolution is modified in this interacting scalar field-radiation model, we employ the standard fitting formula for $z_{\text{drag}}$
as an effective approximation, which is justified by the small values of $|\epsilon|$ allowed by the data. In this sense, full Boltzmann codes are not required at the background level.

In addition, we include nine new measurements from the DESI 2025 DR2 release \cite{desi2025} with redshift range of $0.295<z<2.330$, with eight values of the comoving distance ratio \(D_M/r_d\), and one of the dilation scale ratio \(D_V/r_d\), where \(r_d = r_s(z_{\text{drag}})\). The comoving distance \(D_M(z)\) and the dilation scale \(D_V(z)\) \citep{DILATIONSCALE2005} are defined as follows:
\begin{align}
D_M(z) &= \frac{c}{H_0} \int_0^z \frac{dz'}{E(z')}, \\
D_H(z) &= \frac{c}{H(z)}, \\
D_V(z) &= \left[z D_H(z) D_M^2(z)\right]^{1/3},
\end{align}
with normalized observables
\begin{equation}
D_{M,rd} = \frac{D_M}{r_d}, \quad D_{V,rd} = \frac{D_V}{r_d}.
\end{equation}

The contribution from DESI is encoded in the chi-square function
\begin{equation}
\chi^2_{D_V,D_M} = \sum_{i} \frac{\left(D_i^{\text{obs}} - D_i^{\text{th}}(z_i)\right)^2}{\sigma_{D_i}^2},
\end{equation}
where \(D_i = \{ D_{M,rd}, D_{V,rd} \}\), and \(\sigma_{D_i}\) are the reported uncertainties.

The total BAO chi-square is then
\begin{equation}
\chi^2_{\text{BAO}} = \chi^2_{\theta} + \chi^2_{D_V,D_M}.
\end{equation}

BAO measurements are particularly sensitive to the interacting model through the modified sound horizon, making them a key dataset for constraining $\epsilon$.
\subsection{Cosmic Microwave Background}

In the early universe, photons and baryons formed a tightly coupled plasma where radiation pressure and gravity drove acoustic oscillations. As the universe expanded and cooled, electrons and protons recombined into neutral hydrogen at redshift $z_\ast \sim 1100$, sharply reducing the Thompson scattering rate. Since this epoch, photons traveled freely through the universe and are observed today as the CMB. The pattern of temperature anisotropies in the CMB encodes precise information about the geometry and composition of the universe and is therefore a powerful probe to estimate the cosmological parameters with high precision.

We use the compressed \emph{distance posteriors} extracted from the positions of the acoustic peaks. These quantities are the acoustic scale $l_A$, the shift parameter $R$, and the decoupling redshift $z_\ast$, these are known to be almost independent of the specific dark energy model and can thus be safely employed to test the parameters of alternative cosmologies.

The acoustic scale is defined as
\begin{equation}
l_A \;=\; \pi\,\frac{r(z_\ast)}{r_s(z_\ast)}\,,
\end{equation}
where $r_s$ is the sound horizon \ref{eq:24} at the redshift of decoupling $z_\ast$ given by Hu and Sugiyama~\cite{Hu_1996},
\begin{equation}
z_\ast \;=\; 1048\,
\big[1 + 0.00124\,(\Omega_{b0}h^2)^{-0.738}\big]\,
\big[1 + g_1\,(\Omega_{m0}h^2)^{g_2}\big]\,,
\end{equation}
with
\begin{equation}
g_1 \;=\; \frac{0.0783\,(\Omega_{b0}h^2)^{-0.238}}{1 + 39.5\,(\Omega_{b0}h^2)^{0.763}}\,,
\qquad
g_2 \;=\; \frac{0.560}{1 + 21.1\,(\Omega_{b0}h^2)^{1.81}}\,.
\end{equation}

The shift parameter is defined as \cite{SHIFT_PARAMETER}
\begin{equation}
R \;=\; \frac{\sqrt{\Omega_{m0}}\,H_0}{c}\; r(z_\ast)\,,
\end{equation}
where $\Omega_{m0}$ includes baryon and dark matter.

Thus, the $\chi^2$ for the CMB data is constructed as
\begin{equation}\tag{28}
\chi^2_{\rm CMB} \;=\; X^{\mathrm T}\,{\rm Cov}^{-1}_{\rm CMB}\,X\,,
\end{equation}
where ${\rm Cov}^{-1}_{\rm CMB}$ is the inverse covariance matrix of the distance posteriors and
\begin{equation}
X \;=\;
\begin{pmatrix}
l_A^{\rm th} - l_A^{\rm obs} \\
R^{\rm th}   - R^{\rm obs} \\
z_\ast^{\rm th} - z_\ast^{\rm obs}
\end{pmatrix},
\end{equation}
with the superscripts ${\rm th}$ and ${\rm obs}$ referring to the theoretical and observational estimates, respectively.

\subsection{Comparison of models} \label{comparison}
After obtaining the constraints from the different cosmological data, we compare the different DE models using the Akaike information criterion (AIC) \cite{1974ITAC...19..716A} and Bayesian information criterion (BIC) \cite{BIC1974}, defined as:

\begin{equation}
\text{AIC} = \chi^2_{\text{min}} + 2\alpha,
\end{equation}

\begin{equation}
\text{BIC} = \chi^2_{\text{min}} + \alpha \ln N,
\end{equation}
where $\chi^2_{\text{min}}$ is the chi-square obtained from the best fit of the parameters, $\alpha$ is the number of parameters, and $N$ is the number of data points used in the fit. The methodology consists of computing the difference between the value of each information criterion and that of a reference model, defined as the one with the minimum AIC or BIC. Specifically, $\Delta \text{AIC} = \text{AIC}i - \text{AIC}{\text{min}}$ provides insight into the relative support of the models, whilst $\Delta \text{BIC} = \text{BIC}i - \text{BIC}{\text{min}}$ quantifies the strength of evidence against a given model, as summarized in Table \ref{fig: AIC BIC}. 

\begin{table}\label{table1}
\centering
\begin{tabular}{|c|c|}
\hline
$\Delta$AIC & Empirical support for model $i$ \\
\hline
0 - 2 & Substantial \\
4 - 7 & Considerably less \\
$>$ 10 & Essentially none \\
\hline
$\Delta$BIC & Evidence against model $i$ \\
\hline
0 - 2 & Not worth more than a bare mention \\
2 - 6 & Positive \\
6 - 10 & Strong \\
$>$ 10 & Very strong \\
\hline
\end{tabular}
\caption{Reference values for the $\Delta$AIC and $\Delta$BIC criteria. Note that smaller values of $\Delta$AIC and $\Delta$BIC indicate a preference for the model.}
\label{fig: AIC BIC}
\end{table}
\section{Results}\label{Results}

In this section we present the main results for the constraints on the interacting scalar–radiation model obtained from individual datasets and their joint combination. We consider both a fully free prior on $\Omega_{m0}$ (\textbf{CI}) and a Planck-motivated Gaussian prior (\textbf{CII}), adopting flat priors for the remaining parameters. All other cosmological quantities are fixed to their standard values. First, it is discussed the posterior distributions of the model parameters, followed by the reconstructed expansion history and kinematic diagnostics, and conclude with a statistical comparison to $\Lambda$CDM model.

\subsection{Bayesian analysis results}
\begin{figure}
    \centering
    \includegraphics[width=1.0\linewidth]{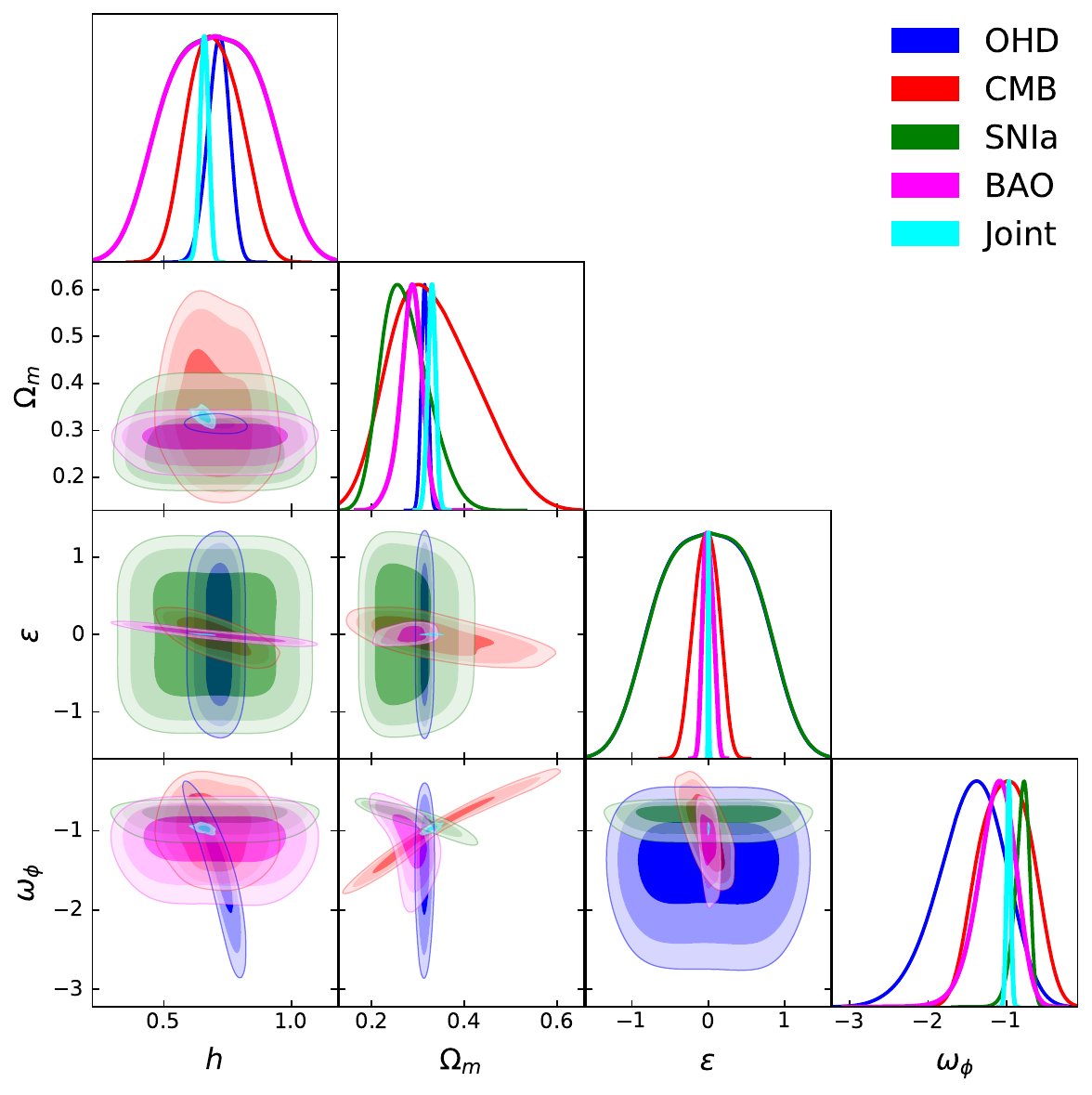}
    \caption{1D posterior distributions and 2D confidence contours for the free parameters characterizing the case \textbf{CI} of the interacting scalar–radiation model, using OHD, CMB, SNIa, BAO, and their Joint combination.}
    \label{fig:contours1}
\end{figure}

\begin{figure}
    \centering
    \includegraphics[width=1.0\linewidth]{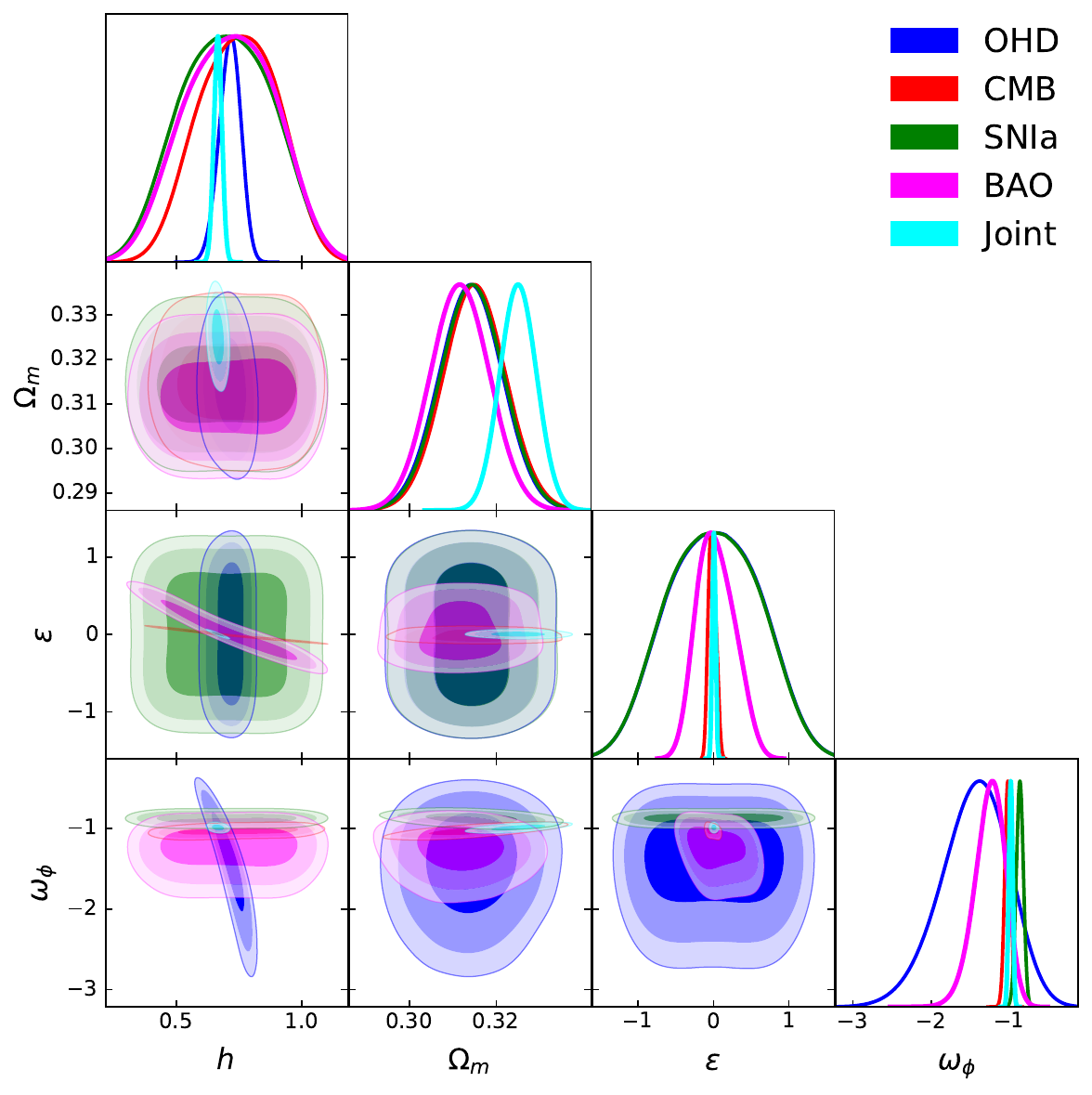}
    \caption{1D posterior distributions and 2D confidence contours for the free parameters characterizing the case \textbf{CII} of the interacting scalar–radiation model, using OHD, CMB, SNIa, BAO, and their Joint combination.}
    \label{fig:contours2}
\end{figure}

\begin{figure}[t]
  \centering
  \begin{minipage}[b]{0.5\textwidth}
    \centering
    \includegraphics[width=\linewidth]{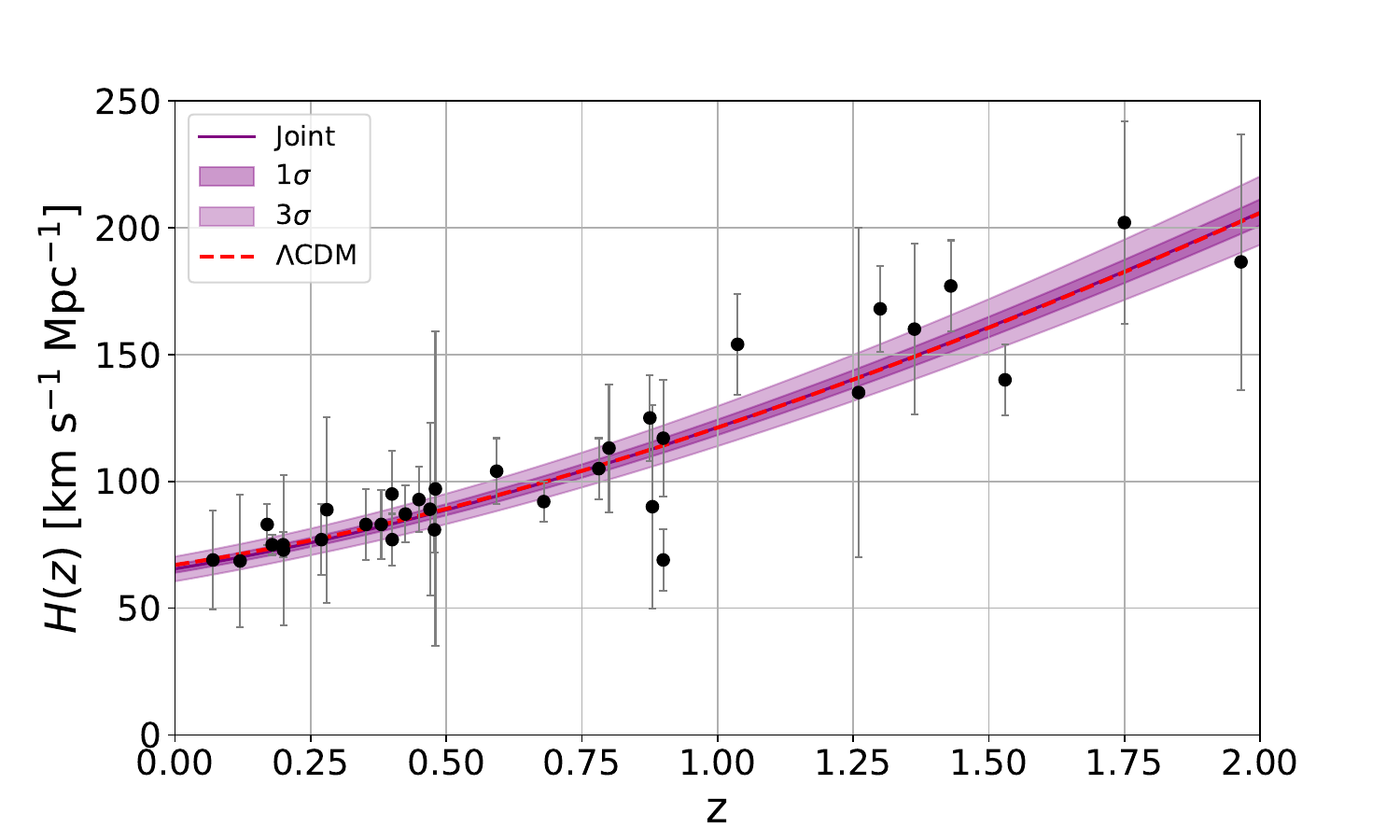}
  \end{minipage}\hfill
  \begin{minipage}[b]{0.5\textwidth}
    \centering
    \includegraphics[width=\linewidth]{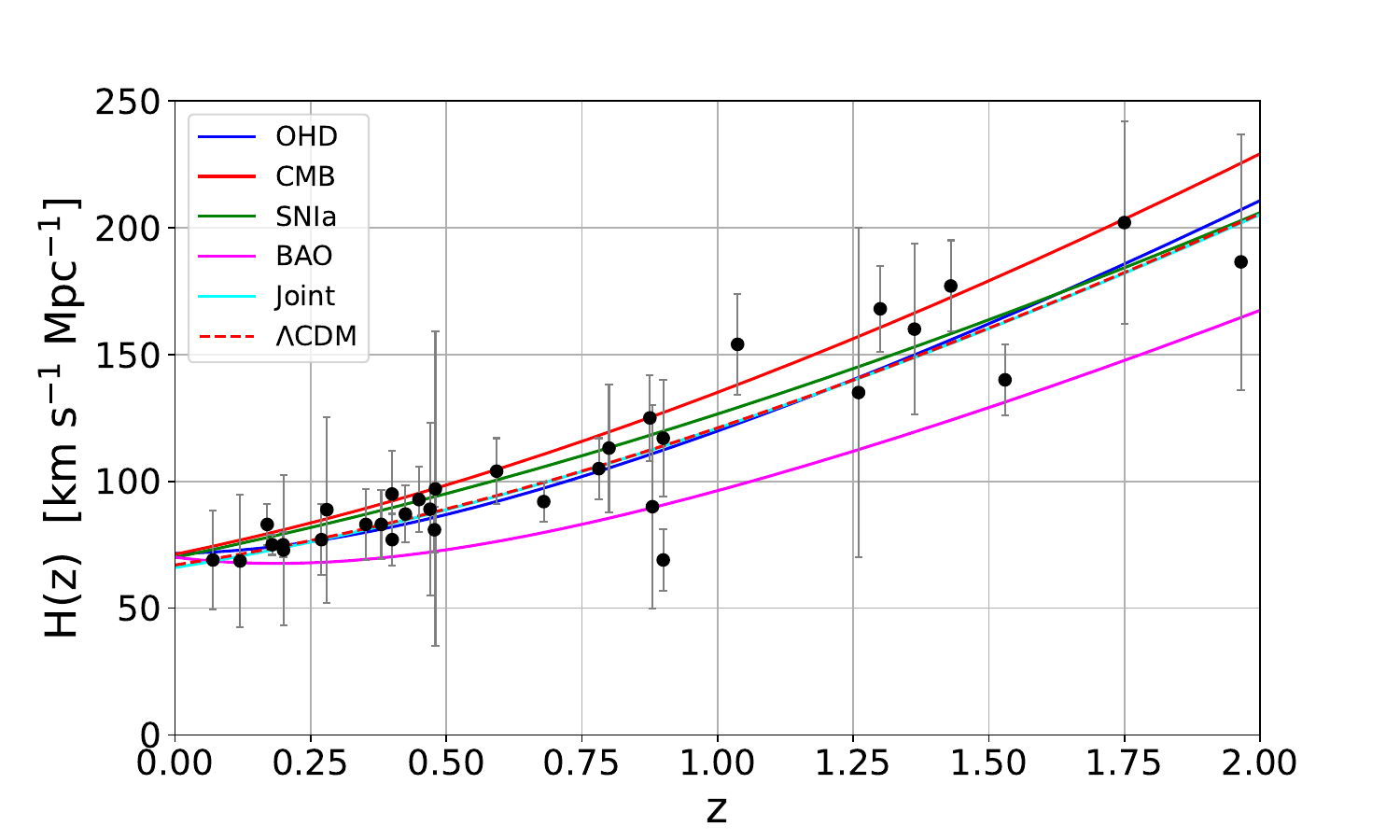}
  \end{minipage}
  \caption{Reconstruction of the Hubble rate $H(z)$ for the interacting scalar–radiation model case \textbf{CI} with the respective data points from OHD data. Left panel: reconstruction from the joint analysis
compared to the best-fit $\Lambda$CDM model, where shaded regions indicate the $1\sigma$
(darker) and $3\sigma$ (lighter) confidence bands and their uncertainties. Right panel: Best-fit constraints from OHD, CMB, SNIa, BAO, and their joint combination.}
  \label{fig:HZ}
\end{figure}

\begin{figure}[t]
  \centering
  \begin{minipage}[b]{0.5\textwidth}
    \centering
    \includegraphics[width=\linewidth]{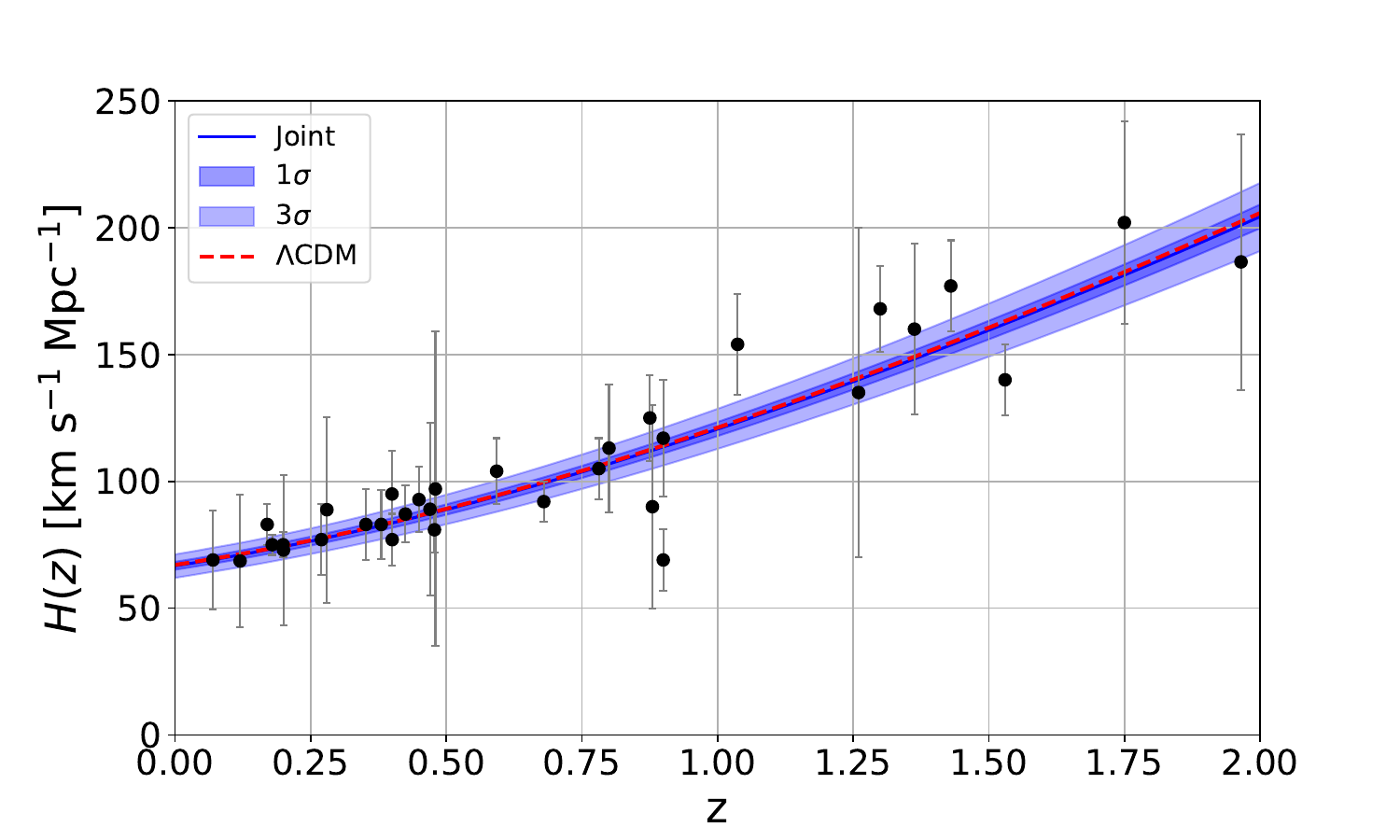}
  \end{minipage}\hfill
  \begin{minipage}[b]{0.5\textwidth}
    \centering
    \includegraphics[width=\linewidth]{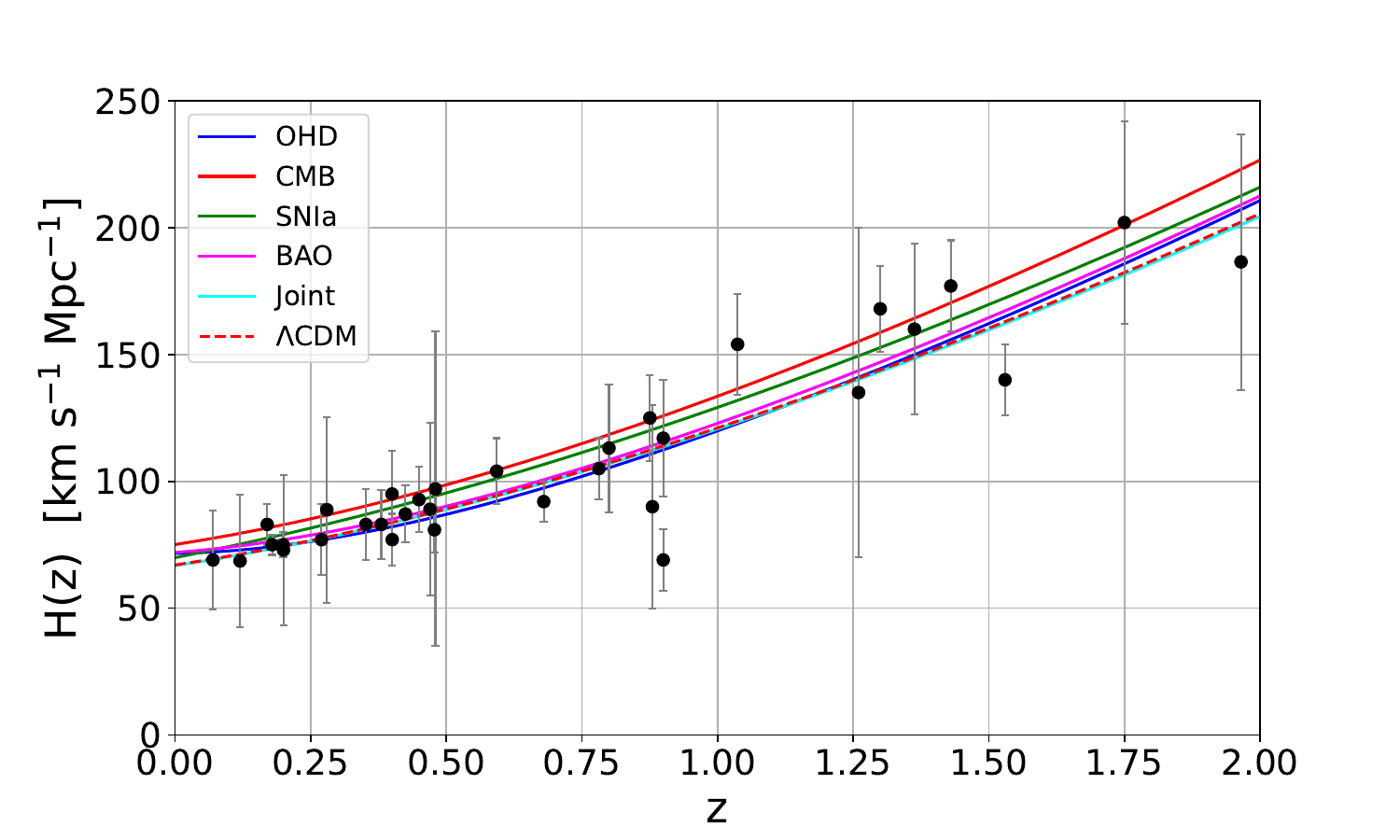}
  \end{minipage}
  \caption{Reconstruction of the Hubble parameter $H(z)$ for the interacting scalar-radiation model case \textbf{CII} with the respective data points from OHD data. Left panel: Best-fit curve from the Joint analysis
compared to the best-fit from the $\Lambda$CDM model, where shaded regions indicate the $1\sigma$
(darker) and $3\sigma$ (lighter) confidence bands and their uncertainties. Right panel: Best-fit curves constraints from OHD, CMB, SNIa, BAO, and their joint combination.}
  \label{fig:HZ_C2}
\end{figure}

We performed a Bayesian analysis using the MCMC python module \verb|emcee| \cite{mcmchammer} setting 400 walkers near the maximum probability region and carried out a burn-in phase until we reached the convergence, after that it is performed 5000 MCMC steps. The convergence of the MCMC analysis is tested using an autocorrelation time test discussed by \cite{Sokal1996MonteCM}. Flat priors were assumed for the parameters $\epsilon$: [-1.0,1.0] and $\omega_{\phi}$: [-3.0,-0.3]. \\
Figs.\ref{fig:contours1} and \ref{fig:contours2} show the marginalized one-dimensional posterior distributions and two-dimensional confidence contours for the model parameters within $1\sigma$, $2\sigma$, and $3\sigma$ (from darker to lighter bands), obtained from OHD, CMB, SNIa, BAO data sets and their Joint combination. A significant result of our analysis is the strong degeneracy between the reduced Hubble constant $h$ and the matter density $\Omega_{m0}$. This degeneracy is clearly shown by the comparison between the \textbf{CI} and \textbf{CII} cases. For \textbf{CI}, where $\Omega_{m0}$ varies freely, the coupling yields a slight upward shift in $h$, but this happens via correlated deviations of $\Omega_{m0}$ from its $\Lambda$CDM value. Conversely, for \textbf{CII} (Planck-motivated prior on $\Omega_{m0}$), this freedom is restricted and significantly limits the range at which $h$ can be shifted. On the other hand, 
The marginalized one-dimensional posterior distributions for the interaction parameter $\epsilon$ are centered near zero in both the \textbf{CI} and \textbf{CII} cases, indicating that the current background data do not require a nonzero interaction but allow minor deviations from standard radiation scaling. The two-dimensional confidence contours reveal that $\epsilon$ is strongly degenerate 
with $h$ and $\Omega_{m0}$, with negative values of $\epsilon$ correlated with higher inferred $h$ in \textbf{CI} case. In contrast, this degeneracy becomes significantly restricted when a Planck-motivated prior on $\Omega_{m0}$ is imposed in the \textbf{CII} case. In this sense, 
different dataset combinations exhibit mild preferences for opposite signs of the interaction, highlighting the sensitivity of early-Universe observables to the direction of energy exchange.
shifts ($z\gtrsim 1$), where the coupling modifies the early expansion rate through its effect on the radiation sector. The Hubble parameter reconstruction is shown in figs. \ref{fig:HZ} and \ref{fig:HZ_C2} for the CI and CII cases, respectively. 
In both scenarios, the expansion history remains consistent with the $\Lambda$CDM prediction to 1$\sigma$ within the inferred 
confidence regions, with only mild deviations driven by parameter degeneracies related to the interaction sector.

\begin{figure}[t]
  \centering
  \begin{minipage}[b]{0.5\textwidth}
    \centering
    \includegraphics[width=\linewidth]{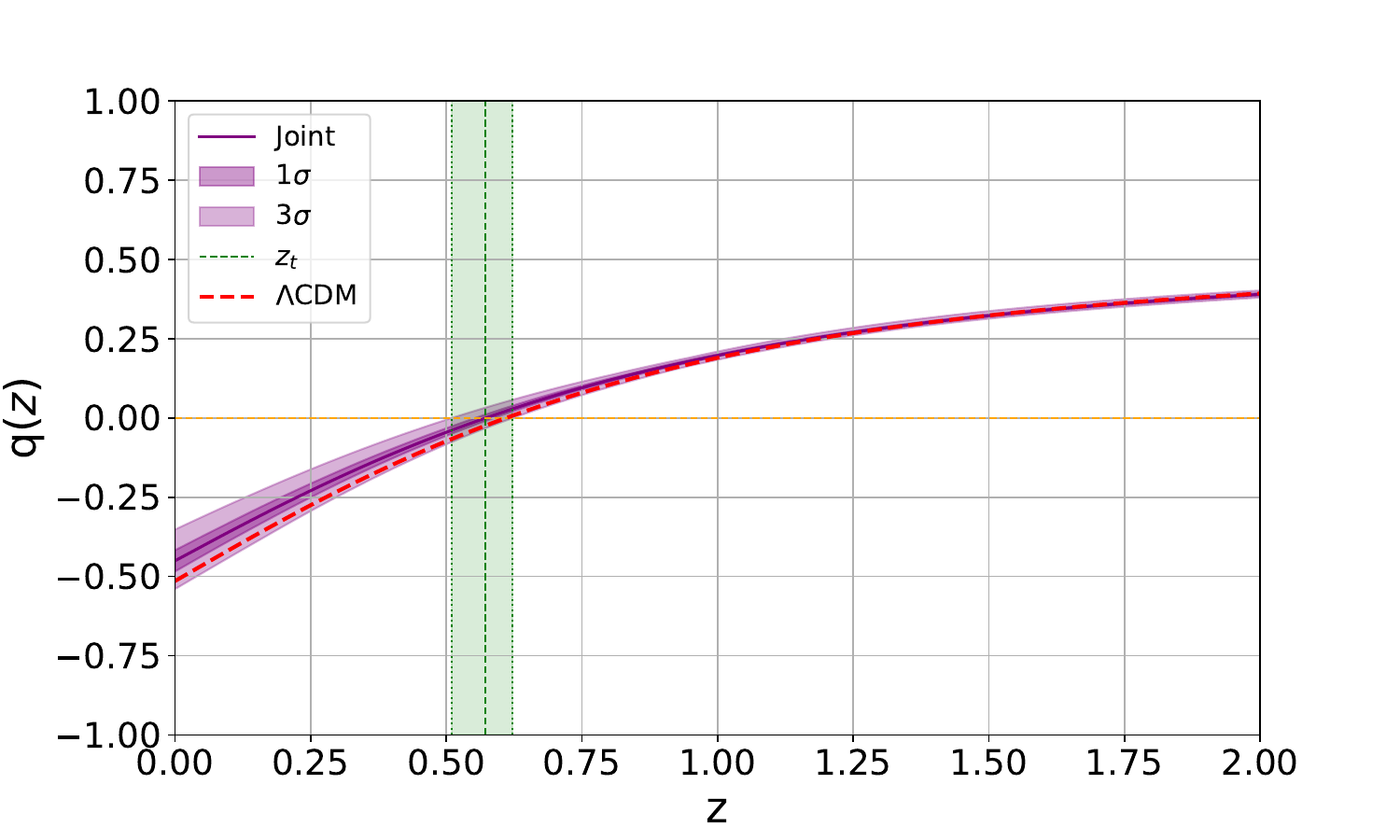}
  \end{minipage}\hfill
  \begin{minipage}[b]{0.5\textwidth}
    \centering
    \includegraphics[width=\linewidth]{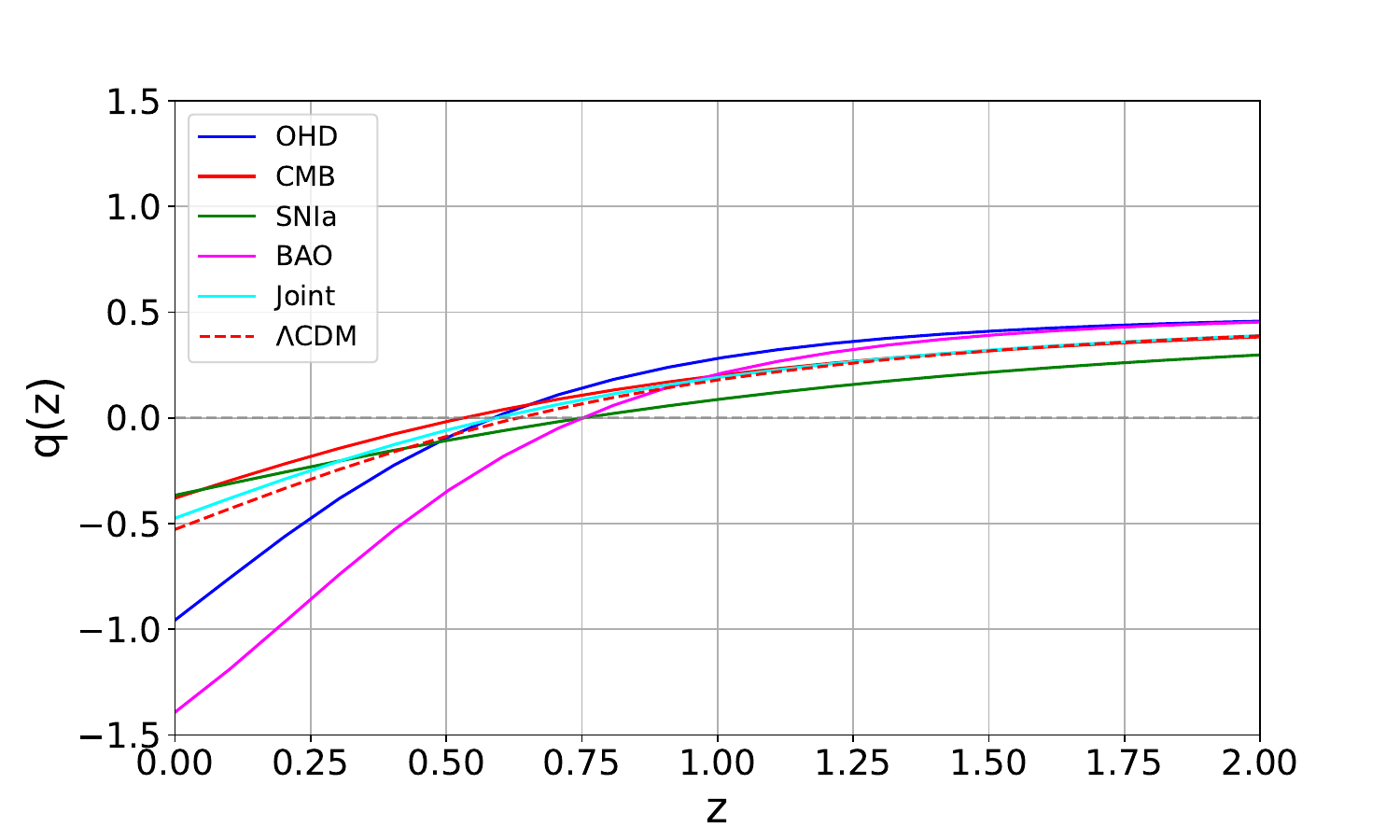}
  \end{minipage}
  \caption{Reconstruction of the deceleration parameter $q(z)$ for the interacting scalar-radiation model case \textbf{CI}
Left panel: Best-fit curve from the Joint analysis compared to the best-fit from the $\Lambda$CDM model, where shaded regions indicate the $1\sigma$ (darker) and $3\sigma$ (lighter) confidence bands. 
Right panel: Best-fit constraints from OHD, CMB, SNIa, BAO and their Joint combination.}
  \label{fig:qz_C1}
\end{figure}

\begin{figure}[t]
  \centering
  \begin{minipage}[b]{0.5\textwidth}
    \centering
    \includegraphics[width=\linewidth]{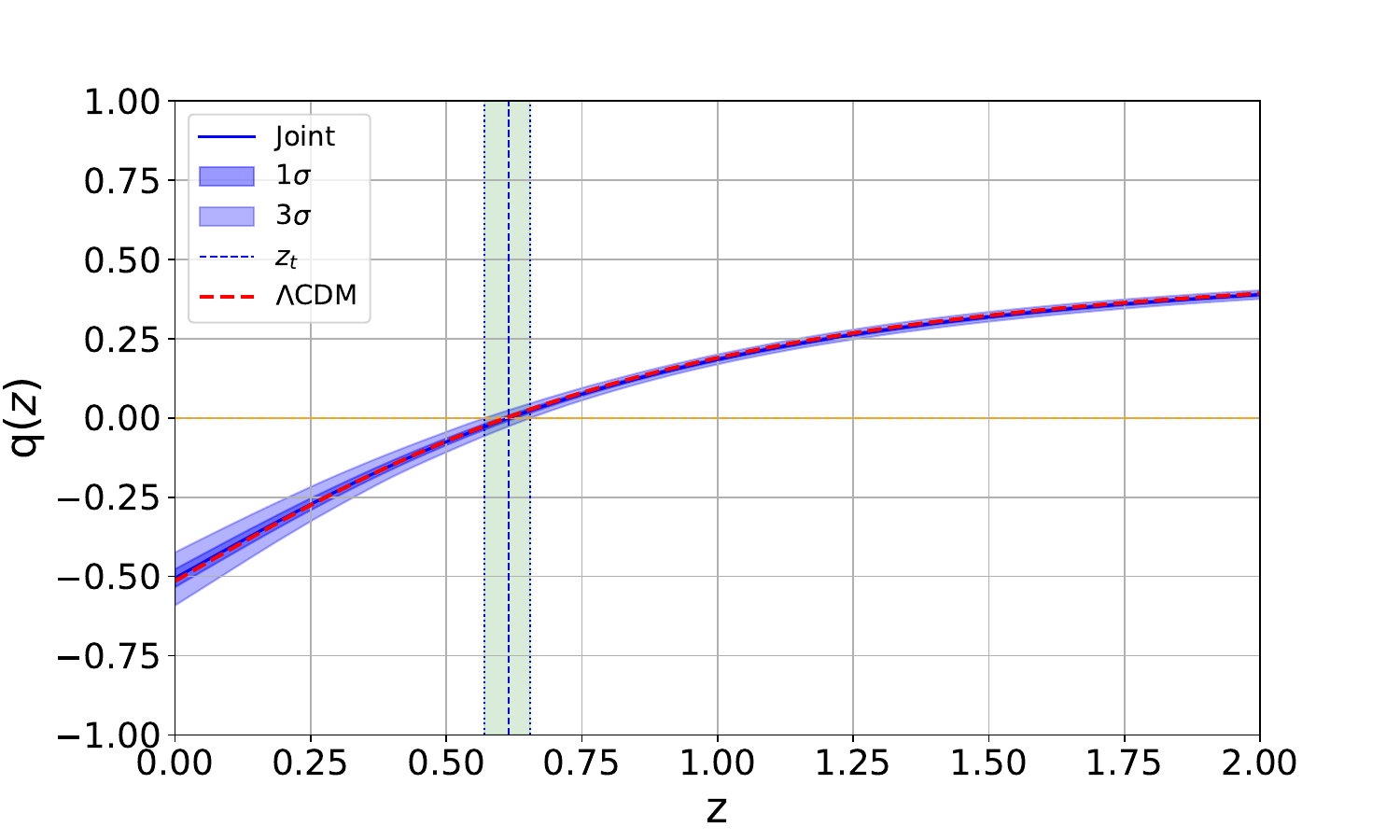}
  \end{minipage}\hfill
  \begin{minipage}[b]{0.5\textwidth}
    \centering
    \includegraphics[width=\linewidth]{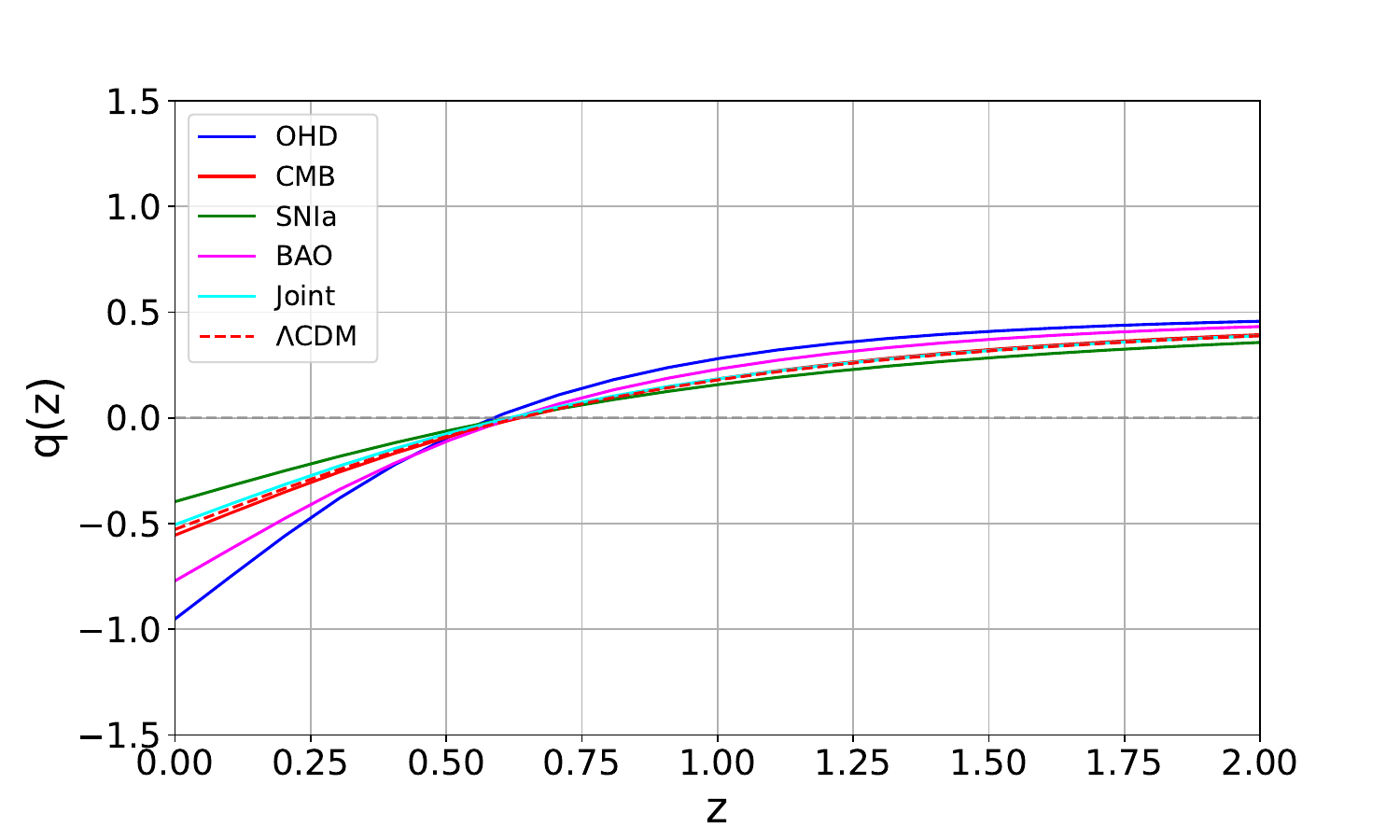}
  \end{minipage}
  \caption{Reconstruction of the deceleration parameter $q(z)$ for the interacting scalar-radiation model case \textbf{CII} 
Left panel: Best-fit curve from the Joint analysis compared to the best-fit from the $\Lambda$CDM model, where shaded regions indicate the $1\sigma$ (darker) and $3\sigma$ (lighter) confidence bands. 
Right panel: Best-fit constraints from OHD, CMB, SNIa, BAO and their Joint combination.}
  \label{fig:qz_C2}
\end{figure}

The deceleration parameter $q(z)$ is shown in Figs. 
\ref{fig:qz_C1} and 
\ref{fig:qz_C2}, illustrating a shift from deceleration to acceleration at redshifts aligning with $\Lambda$CDM expectations. In both \textbf{CI} and \textbf{CII}, the current value $q(z=0)$ remains close to the standard model prediction, with slight deviations attributable to underlying parameter degeneracies rather than indicating a fundamentally different acceleration mechanism. The estimated transition redshift $z_t$ aligns with $\Lambda$CDM within the margin of error.

\begin{figure}[t]
  \centering
  \begin{minipage}[b]{0.5\textwidth}
    \centering
    \includegraphics[width=\linewidth]{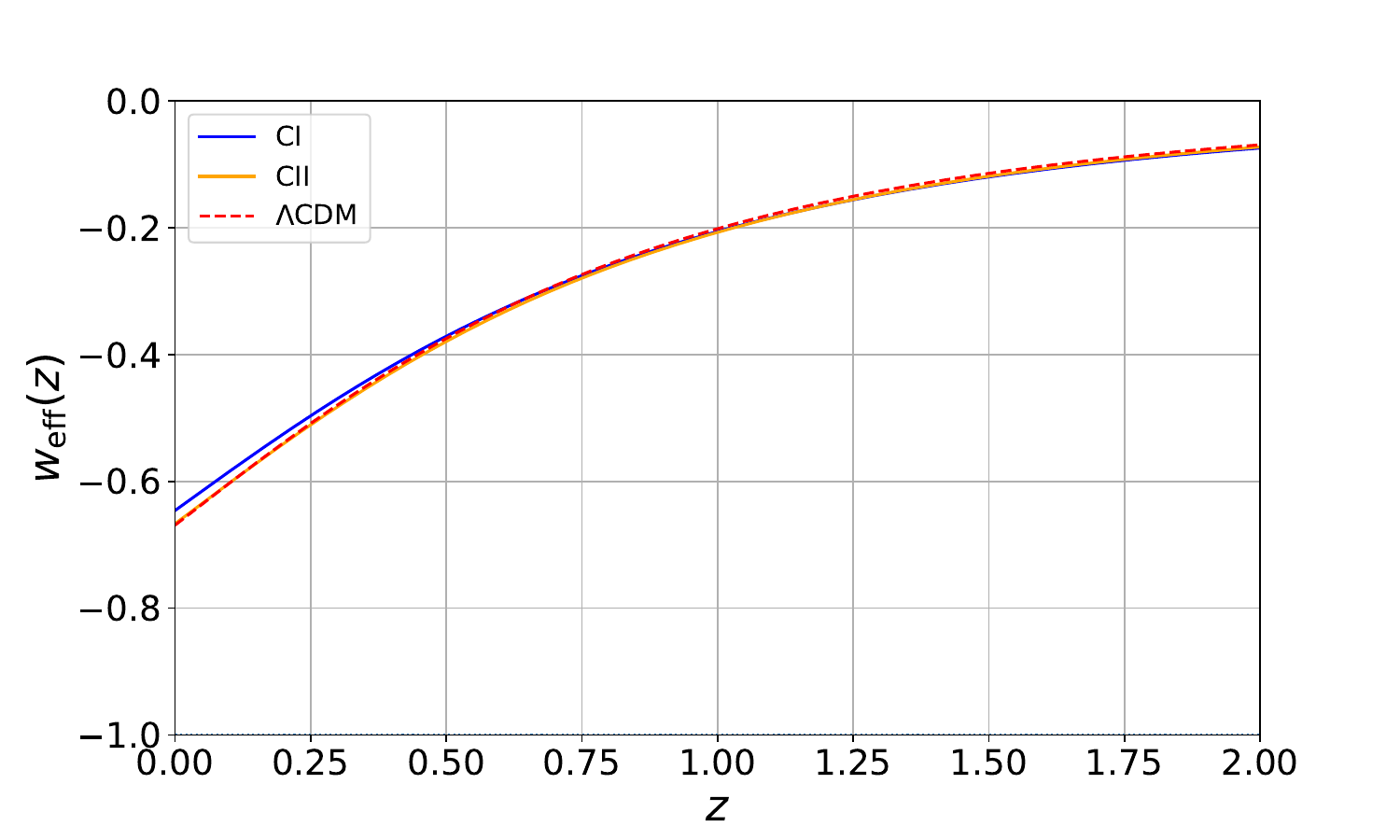}
  \end{minipage}\hfill
  \begin{minipage}[b]{0.5\textwidth}
    \centering
    \includegraphics[width=\linewidth]{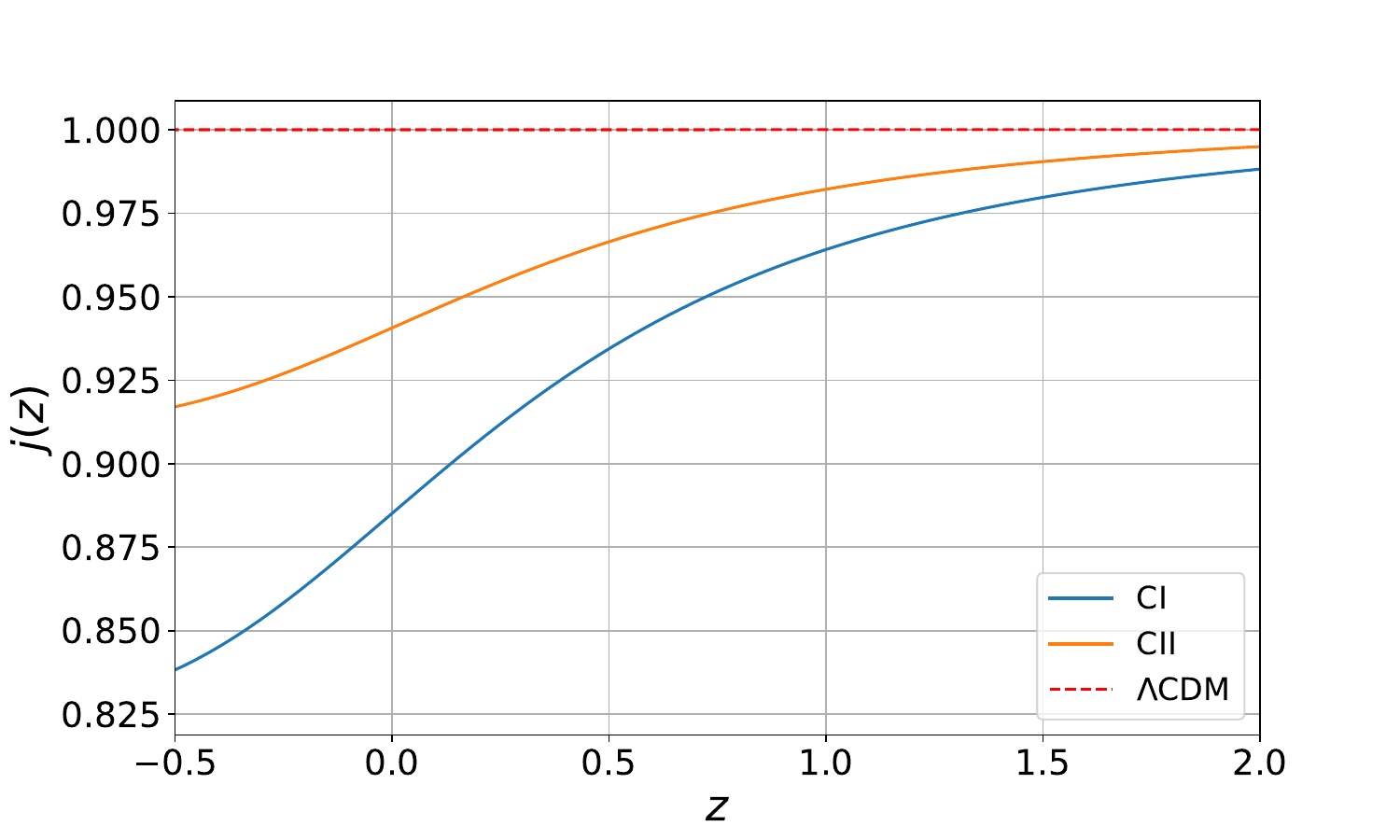}
  \end{minipage}
  \caption{Left panel: $w_{\rm}(z)$ best-fit curve from the Joint analysis for both cases \textbf{CI} and \textbf{CII} of the interacting scalar-radiation model. Right panel: Reconstruction of the Jerk parameter $j(z)$ from the Joint anaylisis for both cases \textbf{CI} and \textbf{CII} of the interacting scalar-radiation model. The $\Lambda$CDM prediction is shown for comparison for the left panel and right panel.}
  \label{fig:EoS_jerk}
\end{figure}

The left panel in Fig.\ref{fig:EoS_jerk} shows the reconstructed effective EoS 
$w_{eff}$, together with the scalar-field and radiation contributions. The interacting scalar field behaves as a quintessence-like component at late times, with 
$w_{\phi}$ remaining close to $-1$, while deviations at earlier epochs reflect the interaction with radiation. These departures remain modest and do not produce dramatic changes in the late-time acceleration history.

\begin{figure}[t]
  \centering
  \begin{minipage}[b]{0.50\textwidth}
    \centering
    \includegraphics[width=\linewidth]{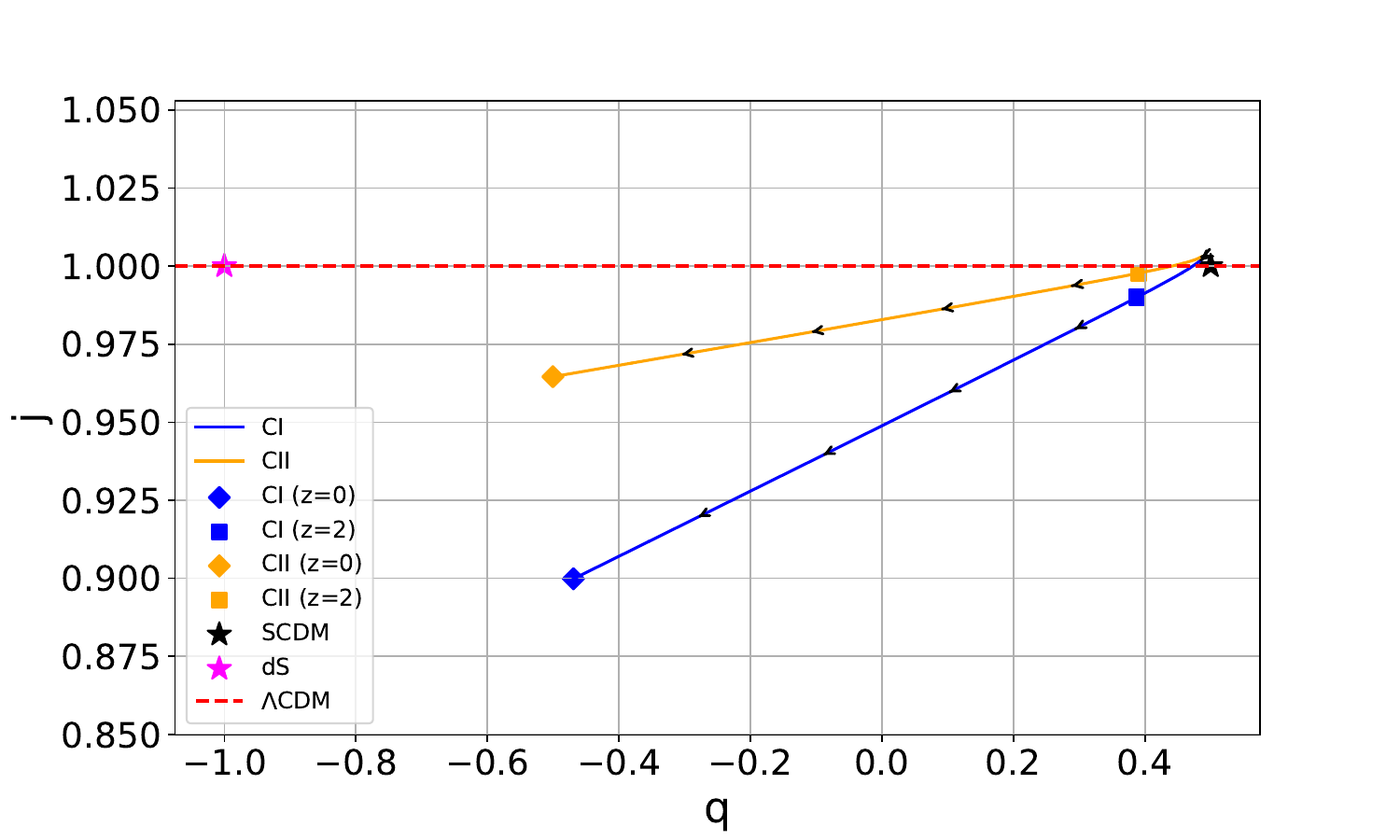}
  \end{minipage}\hfill
  \begin{minipage}[b]{0.50\textwidth}
    \centering
    \includegraphics[width=\linewidth]{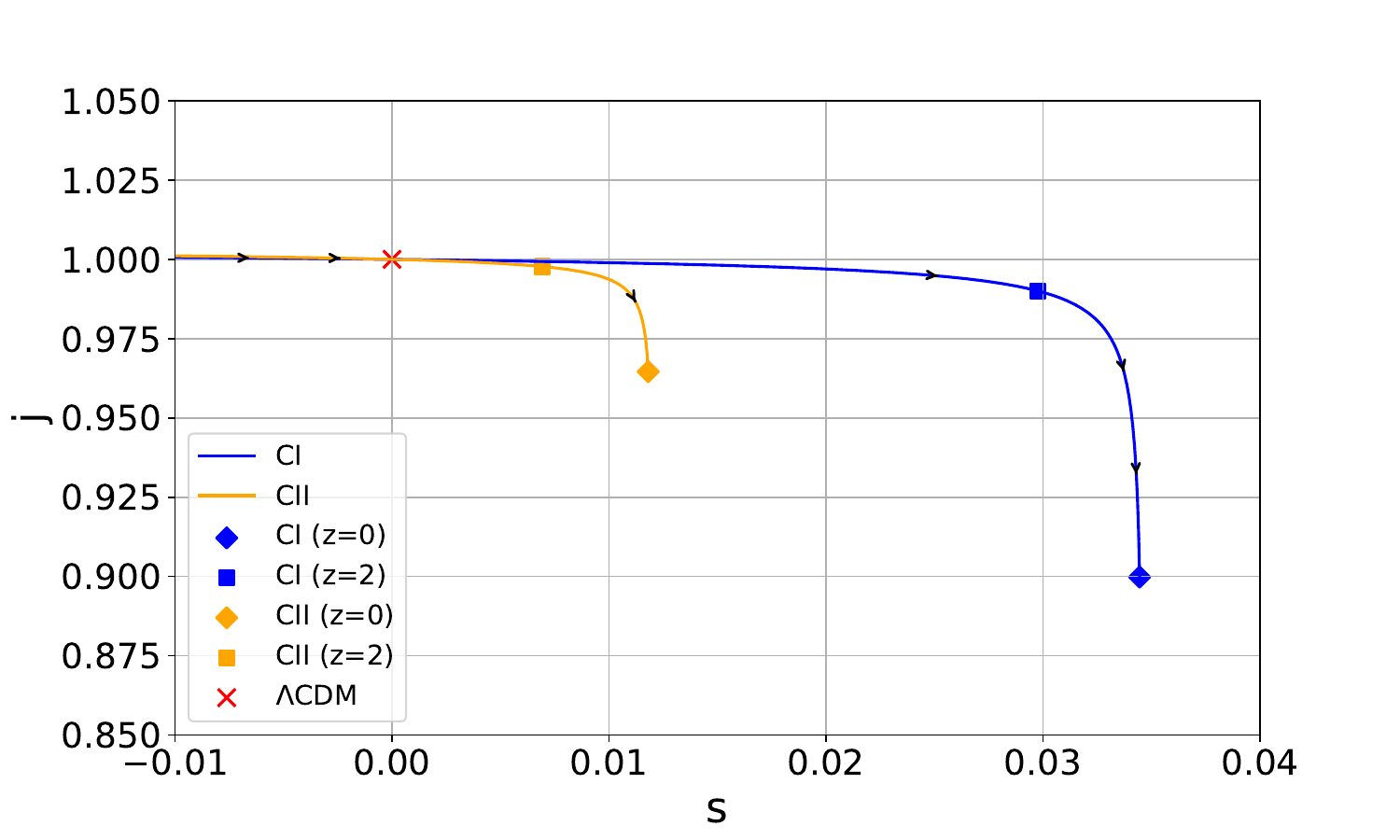}
  \end{minipage}
  \caption{Statefinder diagnostics with the Joint best-fit values for IRSF models CI and CII compared with $\Lambda$CDM. 
The curves show the evolution from high to low redshift (arrows point toward decreasing $z$). 
Open markers indicate snapshots at $z=2$ (open square) and $z=0$ (open diamond).
\textbf{Left panel}: $(q,j)$ plane shows the dS fixed point is at $(q,j)=(-1,1)$ (filled diamond), 
which acts as the late time attractor for $\Lambda$CDM. Present-day locations for IRSF CI and IRSF CII are highlighted with open diamonds.
\textbf{Right panel}: $(s,j)$ plane shows the fixed points as filled markers: $\Lambda$CDM at $s,j=(0,1)$ (filled star) and SCDM at $(1,1)$ (filled square). }
  \label{fig:statefinder}
\end{figure}

The jerk parameter $j(z)$ and the statefinder diagnostics, shown in Figures \ref{fig:EoS_jerk} and \ref{fig:statefinder}, provide complementary kinematic characterization of the model. In the $(q,j)$ and $(s,j)$ planes, both \textbf{CI} and \textbf{CII} trajectories approach the $\Lambda$CDM fixed point $(j=1)$, given by Planck \cite{Planck:2020}, at late times, indicating that the present-day expansion remains close to the standard cosmological attractor. Minor deviations at higher redshifts reflect the effects of the interaction and modified radiation scaling but do not signal qualitatively distinct late-time behavior.

Table \ref{tab:tabla2} presents the mean values of the constraints on our model for cases \textbf{CI} and \textbf{CII}, obtained from the OHD, CMB, SNIa, and BAO data sets and their joint combination at the 3$\sigma$ level, along with the $\chi^2$ and the reduced chi-squared $\chi^2_{\mathrm{red}}$. For the $\Lambda$CDM model, the following set of mean parameter values is obtained at 3$\sigma$: $H(z=0)=67.010^{+0.006}_{-0.006}\,\mathrm{km\,s^{-1}\,Mpc^{-1}}$ $q(z=0)=-0.511^{+0.016}_{-0.015}$, $z_{t}=0.606^{+0.026}_{-0.024}$, and $t(z=0)=13.764^{+0.025}_{-0.025}\,[\mathrm{Gyr}]$.
For case \textbf{CI}, we find: $H(z=0)=65.49^{+1.94}_{-1.87}\,\mathrm{km\,s^{-1}\,Mpc^{-1}}$, $q(z=0)=-0.451^{+0.033}_{-0.034}$, $j(z=0)=0.90971^{+0.0823}_{0.06635}$, $z_{t}=0.573^{+0.050}_{-0.062}$ and $t(z=0)=13.843^{+0.932}_{-0.918}\ \text{{[Gyr]}}$. For case \textbf{CII}, the corresponding values are: $H(z=0)=66.6799^{+1.600}_{-1.529}\,\mathrm{km\,s^{-1}\,Mpc^{-1}}$, $q(z=0)=-0.5051^{+0.080}_{-0.086}$, $j(z=0)=0.96447^{+0.0707}_{0.0660}$, $z_{t}=0.6146^{+0.0400}_{-0.0440}$ and $t(z=0)=13.836^{+0.970}_{-0.818}\ \text{{[Gyr]}}$. 

The AIC and BIC for the interacting model relative to $\Lambda$CDM are summarized in Table \ref{tab:AICBIC}. While \textbf{CI} shows mild support under AIC, the BIC consistently favors $\Lambda$CDM, indicating that the interacting model remains statistically competitive but not decisively preferred at the background level.

As an additional early-Universe consistency check, we examine whether the
jointly constrained interaction parameter satisfies BBN constraints. 
From the joint analysis, the interaction parameter is constrained to
$|\epsilon| \lesssim \mathcal{O}(10^{-2})$ (95\% C.L.), implying an effective
shift $|\Delta N_{\rm eff}| \lesssim 0.3$. This lies well within current BBN bounds, confirming that the interacting scalar-radiation
model remains fully consistent with constraints on light primordial elements.

\begin{table*}[t]
\centering
\resizebox{\textwidth}{!}{%
\begin{tabular}{|l|c|c|c|c|c|c|}
\hline
\multicolumn{7}{|c|}{\textbf{CI}}\\
\hline
Dataset & $h$ & $\Omega_{m0}$ & $\epsilon$ & $\omega_{\phi}$ & $\chi^2_{\min}$ & $\chi^2_{\rm red}$ \\
\hline
OHD &
$0.75305^{+0.08565}_{-0.07307}$ &
$0.26760^{+0.05464}_{-0.04104}$ &
$0.00750^{+0.67754}_{-0.68364}$ &
$-1.56973^{+0.52066}_{-0.60300}$ &
$27.9030$ & $0.8455$ \\
\hline
CMB  &
$0.712150^{+0.286094}_{-0.214327}$ &
$0.349801^{+0.249005}_{-0.149252}$ &
$-0.020894^{+0.109639}_{-0.131798}$ &
$-0.902020^{+0.592999}_{-0.749240}$ &
$2.0461$ & $.6832$ \\
\hline
BAO &
$0.705433^{+0.293829}_{-0.304734}$ &
$0.286269^{+0.059132}_{-0.083674}$ &
$-0.037276^{+0.637906}_{-0.513885}$ &
$-1.105355^{+0.491861}_{-0.982693}$ &
$243.1090$ & $1.6882$ \\

\hline
SNIa  &
$0.698119^{+0.301030}_{-0.297340}$ &
$0.265149^{+0.172103}_{-0.064969}$ &
$0.014357^{+0.983295}_{-1.011802}$ &
$-0.785907^{+0.157171}_{-0.423696}$ &
1758.212 & 1.0336 \\
\hline
Joint  &
$0.65744^{+0.05079}_{-0.04018}$ &
$0.33068^{+0.03764}_{-0.02285}$ &
$0.00584^{+0.05644}_{-0.05993}$ &
$-0.96549^{+0.10786}_{-0.08710}$ &
2071.523 & 1.1042 \\
\hline
\multicolumn{7}{|c|}{\textbf{CII}}\\
\hline
Dataset & $h$ & $\Omega_{m0}$ & $\epsilon$ & $\omega_{\phi}$ & $\chi^2_{\min}$ & $\chi^2_{\rm red}$ \\
\hline
OHD &
$0.71579^{+0.10441}_{-0.13410}$ &
$0.31408^{+0.02128}_{-0.02087}$ &
$0.00232^{+0.99504}_{-0.99965}$ &
$-1.41647^{+0.93148}_{-1.44134}$ &
$27.9030$ & $0.8455$ \\

\hline
CMB  &
$0.753154^{+0.246069}_{-0.301628}$ &
$0.314942^{+0.020997}_{-0.020492}$ &
$-0.023383^{+0.116560}_{-0.067689}$ &
$-1.025243^{+0.102848}_{-0.122667}$ &
$30.480$ & $2.1602$ \\
\hline
BAO &
$0.719431^{+0.279857}_{-0.318026}$ &
$0.311582^{+0.019780}_{-0.019684}$ &
$-0.001697^{+0.603894}_{-0.406995}$ &
$-1.234899^{+0.413733}_{-0.713242}$ &
$243.1090$ & $1.6882$ \\
\hline

SNIa  &
$0.699110^{+0.300035}_{-0.298366}$ &
$0.314400^{+0.020838}_{-0.020872}$ &
$0.002139^{+0.995441}_{-0.999635}$ &
$-0.871605^{+0.125641}_{-0.137753}$ &
1756.1145 & 1.0941 \\
\hline

Joint  &
$0.664705^{+0.047309}_{-0.046684}$ &
$0.325031^{+0.012404}_{-0.011978}$ &
$0.004708^{+0.064455}_{-0.060595}$ &
$-0.988138^{+0.067459}_{-0.070664}$ &
2070.7781 & 1.1032 \\

\hline
\end{tabular}
}
\caption{Interacting scalar–radiation model cases \textbf{CI} and \textbf{CII}, constrained by OHD, CMB, SNIa, BAO, and their joint combination, with best-fit values and their respective errors within the $3\sigma$ range.}
\label{tab:tabla2}
\end{table*}



\begin{table*}\label{table3}
\centering
\resizebox{.95\linewidth}{!}{%
\begin{tabular}{|c|c|c|c|c|c|c|}
\hline
Data set & \multicolumn{2}{|c|}{$\Lambda$CDM} & \multicolumn{2}{|c|}{CI} & \multicolumn{2}{|c|}{CII} \\
\hline
& $\Delta$ AIC & $\Delta$ BIC & $\Delta$ AIC & $\Delta$BIC & $\Delta$AIC & $\Delta$BIC \\
\hline
Joint & {1.4840} & {0.0} & {0.0} & {5.9878} & {4.64413} & {6.6319} \\
\hline
\end{tabular}}
\caption{Differences in AIC and BIC values with respect to the minimum among the listed models, computed using the chi-square value from the joint analysis.}
\label{tab:AICBIC}
\end{table*}

\section{Discussion and Conclusion}\label{Discussion and Conclusions}

In this work, we have investigated the cosmological consequences of an interacting scalar-radiation model, in which a scalar field is minimally coupled to gravity, but interacts with the radiation sector through a coupling term $C(\phi)=e^{-\sigma \phi}$. This study extends recent work \cite{Bisabr_2025} by incorporating a non-relativistic matter component to provide a more realistic description of the universe's evolution from the radiation-dominated epoch to the present accelerated epoch. 

At the background level, we derived analytical expressions for the expansion history. We characterized the cosmological dynamics using kinematic diagnostics, including the effective equation of state, the deceleration parameter, and statefinder parameters. The interaction modifies the standard radiation dilution law, effectively shifting the sound horizon and modifying early-time expansion while leaving late-time dynamics governed mainly by a quintessence-like scalar field with 
$w_{\phi}\simeq -1$.

Using OHD, SnIa, BAO (including DESI DR2), and compressed CMB distance information, we constrained the model parameters through a Bayesian MCMC analysis. To assess the robustness of the inferred interaction, we considered two complementary cases: (\textbf{CI}) an entirely free prior on $\Omega_{m0}$, and (\textbf{CII}) a Planck-motivated Gaussian prior. This strategy isolates parameter degeneracies and clarifies the role of early-Universe anchoring in determining $H_0$.

Our results indicate that the interaction parameter $\epsilon$ is statistically consistent with zero in both \textbf{CI} and \textbf{CII}, implying that the current background data do not require a non-vanishing scalar-radiation coupling. Nonetheless, small
departures from the standard radiation scaling are allowed and can give rise to non-trivial phenomenology. In particular, negative values of $\epsilon$ reduce the sound horizon, shifting the inferred value of
$H_0$ upward when fitting early-time observables. In the \textbf{CI} case, this mechanism
can partially alleviate the Hubble tension, reducing it to the
$\sim 1\sigma$ level. However, this apparent improvement is accompanied by
correlated shifts in $\Omega_{m0}$, which reach the $\sim 3\sigma$ level
relative to the $\Lambda$CDM reference value when all datasets are combined.

When applying a Planck-motivated prior on $\Omega_{m0}$ (case \textbf{CII}), the ability to adjust $H_0$ is minimal. In this scenario, the inferred expansion history closely aligns with that of the $\Lambda$CDM model, and the residual Hubble tension is only slightly impacted. This comparison emphasizes that an improvement in the $H_0$ discrepancy within this framework is strongly linked to degeneracies with matter density and must be assessed in the context of overall cosmological consistency.

Regarding the reconstructed kinematic quantities, both \textbf{CI} and \textbf{CII} trajectories approach the $\Lambda$CDM fixed point in the statefinder planes as time advances, showing that the current expansion closely aligns with the standard cosmological attractor. Interaction-induced deviations are mainly observed at higher redshifts ($z \gtrsim 1$), where the altered radiation sector becomes more influential. Notably, these deviations are modest and do not give rise to fundamentally different late-time acceleration mechanisms.

Model comparison using information criteria shows that, while the \textbf{CI} case exhibits mild support under the AIC, the BIC consistently favors $\Lambda$CDM due to its smaller number of free parameters. In general, the interacting scalar-radiation scenario is statistically competitive with $\Lambda$CDM at the background level but is not decisively preferred by current data.

Summarizing, the interacting scalar-radiation model constitute a physically
well-motivated and observationally consistent extension of EDE
scenario, which can partially ease the Hubble tension by modifying the early
expansion history; however, this improvement is accompanied by correlated
shifts in other cosmological parameters, most notably $\Omega_{m0}$.
Although current background data do not require the presence of such an
interaction, they allow it at a level that remains fully compatible with
BBN and CMB constraints. A definitive evaluation of the
viability of this class of models will require future studies that incorporate
cosmological perturbations and large-scale structure growth observables, in order to test whether early and late universe tensions can be simultaneously addressed.

\FloatBarrier
\section*{Data availability statement}
All data that support the findings of this study are included within the article (and any supple-
mentary files).
\section*{Acknowledgements}

Nelson Videla acknowledges support from ANID–FONDECYT Grant No.~1220065.

\bibliographystyle{elsarticle-num} 
\bibliography{references}
\end{document}